\documentclass[a4paper,11pt]{article}
\pdfoutput=1 

\usepackage{jheppub} 

\usepackage[UKenglish]{babel}

\usepackage{xcolor}
\usepackage{amsmath}
\usepackage{tensor}
\usepackage[utf8]{inputenc}
\usepackage{graphicx}
\usepackage{array}
\usepackage{tabularx}
\usepackage{amsmath}
\usepackage{amsfonts}
\usepackage{mathrsfs}
\usepackage{slashed}
\usepackage{epigraph}
\usepackage{amssymb}
\usepackage{pifont}
\usepackage{fancyhdr}
\usepackage{amssymb}
\usepackage{graphicx}
\usepackage{longtable}
\usepackage{makecell}
\usepackage{amscd}
\usepackage{booktabs}
\usepackage{bm}
\usepackage{enumerate}

\title{Matching for FCNC effects in the flavour-symmetric SMEFT}
\preprint{\begin{flushright}MITP/18-118\end{flushright}}
\author{Tobias Hurth$^a$,}
\author{Sophie Renner$^a$,}
\author{and William Shepherd$^{a,b}$}

\emailAdd{hurth@uni-mainz.de}
\emailAdd{sorenner@uni-mainz.de}
\emailAdd{shepherd@shsu.edu}

\affiliation{$^a$PRISMA+ Cluster of Excellence \& Mainz Institute of Theoretical Physics, \\ Johannes Gutenberg-Universit\"at Mainz, 55099 Mainz, Germany}
\affiliation{$^b$Physics Department, Sam Houston State University, Huntsville, TX 77431, USA}

\abstract{
We calculate the complete tree and one-loop matching of the dimension-six Standard Model Effective Field Theory (SMEFT) with unbroken $U(3)^5$ flavour symmetry to the operators of the Weak Effective Theory (WET) which are responsible for flavour changing neutral current effects among down-type quarks. We also explicitly calculate the effects of SMEFT corrections to input observables on the WET Wilson coefficients, a necessary step on the way to a well-defined, complete prediction. These results will enable high-precision flavour data to be incorporated into global fits of the SMEFT at high energies, where the flavour symmetry assumption is widespread.
}

\begin{document}

\maketitle

\section{Introduction}

Despite the historically impressive performance of the LHC, with massive datasets delivered at the highest collision energies ever achieved in a laboratory, an understanding of the nature of physics beyond the Standard Model (SM) remains elusive. Since the 2012 discovery of a Higgs-like scalar \cite{Aad:2012tfa,Chatrchyan:2012xdj}  (whose properties continue to look more and more Higgs-like with increasing scrutiny \cite{ATLAS:2018doi,Sirunyan:2018koj}), the SM has been completed, but the questions left unanswered by it remain as compelling as ever.

In the face of this uncertainty as to the nature of whatever new physics (NP) underlies the SM at higher energy scales, it makes sense to remain as agnostic as possible in our interpretations of the data that is available to us. This will allow us to make accurate statements which apply not only to the particular theories which are \emph{en vogue} at the moment, but also to theories which have not yet been dreamt up. In the interest of enabling such an agnostic analysis of physics data, the SM Effective Field Theory (SMEFT) (for a recent review see \cite{Brivio:2017vri}) is an essential tool. This provides the technology for defining a basis of interaction types which, under the assumption that $h(125)$ is the remnant of the electroweak doublet responsible for electroweak symmetry breaking, spans the full set of possibilities that could be induced by NP when probed below its characteristic energy scale, which we denote as $\Lambda$. An analysis framed in terms of the operators of the SMEFT can be straightforwardly mapped into constraints on an arbitrary model of heavy NP using the well-understood technology of amplitude matching, now possible using various automated tools \cite{Criado:2017khh,Bakshi:2018ics}. The interface between different codes, and translation between operator bases, is facilitated by the Wilson coefficient exchange format initiative~\cite{Aebischer:2017ugx}. 

Many steps toward the implementation of SMEFT as a target for analysis have already been taken. The appropriate theoretical underpinnings of the SMEFT theory itself have been developed, notably the determination of a complete basis of operators at dimension-six~\cite{Grzadkowski:2010es} and their renormalization \cite{Jenkins:2013zja,Jenkins:2013wua,Alonso:2013hga}. Precision electroweak data have been used to develop fits of the relevant subset of operators that contribute to those observables \cite{Han:2004az,Berthier:2015oma,Berthier:2015gja,Bjorn:2016zlr,Berthier:2016tkq,Ellis:2018gqa,Almeida:2018cld}.  Loop corrections to very precisely measured and particularly interesting processes have been calculated \cite{Zhang:2013xya,Gauld:2015lmb,Hartmann:2015oia,Gauld:2016kuu,Maltoni:2016yxb,Zhang:2016omx,Hartmann:2016pil,Baglio:2017bfe,Dawson:2018pyl,Dawson:2018jlg,Vryonidou:2018eyv,Dawson:2018liq,Dawson:2018dxp}. LHC searches have also been interpreted in the SMEFT, with techniques to address the unique theoretical errors inherent in high energy searches for EFT effects recently developed~\cite{Alte:2017pme,Alte:2018xgc}.  Ultimately, all of these contributions will need to coalesce into a fully global fit of the SMEFT Wilson coefficients and cutoff scale, as only then will it be possible to truly understand the conservative constraints which can be imposed on an arbitrary model through matching to the SMEFT and comparison of a given parameter point to the likelihood associated with all the relevant measurements.

An important source of precise data which should be used as much as possible is the output of the tremendous effort of the experimental flavour physics community, where measurements of a vast number of processes have been made, nearly all of which indicate that the approximate flavour symmetry which is present in the SM must remain very nearly correct up to scales far higher than we typically expect of NP. 
Many groups have studied the implications of the SMEFT for flavour observables and vice versa (e.g.~\cite{Grzadkowski:2008mf,Lee:2008xs,Drobnak:2011wj,Kamenik:2011dk,Drobnak:2011aa,Alonso:2014csa,Brod:2014hsa,Bobeth:2015zqa,Aebischer:2015fzz,Cirigliano:2016nyn,Gonzalez-Alonso:2016etj,Feruglio:2016gvd,Feruglio:2017rjo,Bordone:2017anc,Bobeth:2017xry,Celis:2017doq,Buttazzo:2017ixm,Cornella:2018tfd,Straub:2018kue,Endo:2018gdn,Silvestrini:2018dos,Descotes-Genon:2018foz,Aebischer:2018iyb}), often focussing on subsets of operators which contribute to particular vertices, or considering explicitly flavour-violating interactions within the SMEFT itself. Codes also exist to perform the running above and below the electroweak scale, and the tree level matching between the SMEFT and the WET~\cite{Celis:2017hod,Aebischer:2018bkb}. 

Here, we tackle the problem with a symmetry-led approach. Taking as a starting point the observation that large flavour violating effects beyond the Standard Model are already ruled out, but retaining the hope of NP at the TeV scale (which could, for instance, address the gauge hierarchy problem), we begin from the simple assumption of an exact flavour symmetry (a particularly strong case of Minimal Flavour Violation~\cite{DAmbrosio:2002vsn}), and include in our theory all operators which are invariant under this.
In this article we present the full tree and one-loop matching between the CP conserving, $U(3)^5$-symmetric SMEFT at dimension-six and down type flavour-changing neutral current operators in the weak effective theory (WET) which arises when the heavy gauge bosons, Higgs boson, and top quark are integrated out of the theory. Flavour violation is due purely to SM effects in the CKM matrix describing the interactions of $W^\pm$ bosons with the quarks. 

These calculations of the loop effects of flavour conserving operators may be relevant even in cases where this symmetry is not imposed {\it ab initio}, when the tree-level flavour changing effects are required to be sufficiently suppressed that loop-level contributions from flavour-symmetric or flavourless couplings become comparable to them. Furthermore, most global fits to SMEFT coefficients using LEP and LHC data have been performed assuming this flavour-symmetric paradigm, so exploring this parameter region in detail in the flavour sector will allow for additional observables to be included in these fits, leading to new and/or tighter constraints. Finally, this assumption represents a ``worst-case scenario'' for flavour searches in the context of roughly TeV-scale NP, so this calculation will give an insight into the smallest effects we should reasonably expect to see if NP is near the TeV scale.

An additional important feature of this article is its inclusion of the non-trivial effects of the SMEFT on observables used as inputs to define Lagrangian parameters \cite{Berthier:2015oma,Brivio:2017bnu}.  In order to make a physical prediction of an observable in quantum field theory it is necessary to define all of the Lagrangian parameters of the theory in terms of observables. In the SM these definitions are so long standing, and the observables used so standardized, that we have grown used to simply assigning numerical values to the Lagrangian parameters as though they were measured themselves, but this is not the case. The SMEFT is not turned on only for ``signal'' processes and inactive in ``input'' measurements; its effects on both measurements must be considered in order to properly predict the sensitivity of any observable to the Wilson coefficients parameterizing new physics effects.

In the next section, we shall discuss the particular set of interactions that arise in the flavour-symmetric limit of the SMEFT. In Sec.~\ref{sec:methods} we lay out our methods and explain how we fix free parameters in the SMEFT Lagrangian using measurements of input parameters. Then, in Sec.~\ref{sec:results} we present the results of our matching calculation between the SMEFT and WET, considering in turn direct contributions from new coupling structures in the SMEFT not present in the SM and contributions from SMEFT effects on the extraction of would-be SM couplings from experimental input measurements. We conclude in Sec.~\ref{sec:conc} with a discussion of the implications and utility of this matching calculation, as well as a mention of future directions. The input dependence of our results is explained in App.~\ref{sec:inputsappendix}, where we provide a separation of the calculation into pieces that are independent of the input parameters chosen, and pieces that arise purely due to SMEFT effects in input parameter measurements, given in two different input schemes.

\section{Flavour-Symmetric SMEFT}
\label{sec:flavoursymm}

\begin{table}
\begin{center}
\begin{tabular}{c c c c c}
\toprule
Group & Operators & $d_i \to d_j \gamma$ &  $d_i \to d_j \,l^+ l^-$ & Meson mixing \\
\midrule
1 & $Q_G$ & -& - & - \\
& $Q_W$ & \ding{52} & \ding{52}&- \\
\midrule
2 & $Q_H$& -& - & -  \\
\midrule
3 & $Q_{H\Box}$& - & -& - \\
& $Q_{HD}$& \ding{52}& \ding{52} &-  \\
\midrule
4 & $Q_{HG}$ & -& - & - \\
& $Q_{HW}$ & -& - & - \\
& $Q_{HB}$ & -& - & -  \\
& $Q_{HWB}$ & \ding{52}& \ding{52} & - \\
\midrule
7 & $Q^{(1)}_{H\ell}$ & - & \ding{52}& -  \\
& $Q^{(3)}_{H\ell}$ & \ding{52}* & \ding{52} & \ding{52}*  \\
& $Q_{He}$ & -& \ding{52} & - \\
& $Q^{(1)}_{Hq}$ & -& \ding{52} & - \\
& $Q^{(3)}_{Hq}$ & \ding{52}& \ding{52} & \ding{52} \\
& $Q_{Hu}$ & -& \ding{52} & -  \\
& $Q_{Hd}$ & -& - & -  \\
\bottomrule
\end{tabular}
\end{center}
\caption{All operators with 2 or fewer fermions that are invariant under CP and the $U(3)^5$ flavour symmetry. Ticks indicate that they contribute to the FCNC processes we consider. An asterisk $(*)$ signifies that the contribution is only indirect, via effects in input parameter measurements.}
\label{table:2fermion}
\end{table}

\begin{table}
\begin{center}
\begin{tabular}{c c c c c}
\toprule
Group & Operators & $d_i \to d_j \gamma$ &  $d_i \to d_j \,l^+ l^-$ & Meson mixing \\
\midrule
8: $(\bar{L}L)(\bar{L}L)$ & $Q_{\ell \ell}$ & \ding{52}* & \ding{52}* & \ding{52}* \\
 & $Q^{(1)}_{q q}$ & -& \ding{52} & \ding{52}  \\
  & $Q^{(3)}_{q q}$ & -& \ding{52} & \ding{52} \\
 & $Q^{(1)}_{\ell q}$& - & \ding{52} & -  \\
  & $Q^{(3)}_{\ell q}$& - & \ding{52} & -  \\
\midrule
8: $(\bar{R}R)(\bar{R}R)$ & $Q_{ee}$& - & - & - \\
& $Q_{uu}$ & -& - & - \\
& $Q_{dd}$ & -& - & - \\
& $Q_{eu}$ & -& \ding{52} & - \\
& $Q_{ed}$ & -& - & - \\
& $Q^{(1)}_{ud}$ & -& - & -  \\
& $Q^{(8)}_{ud}$ & -& - &  -  \\
\midrule
8: $(\bar{L}L)(\bar{R}R)$ & $Q_{\ell e}$ & -& - & -  \\
& $Q_{\ell u}$ & - & \ding{52} & - \\
& $Q_{\ell d}$ & -& - & -\\
& $Q_{qe}$ & -& \ding{52} & - \\
& $Q^{(1)}_{qu}$ & -& - &  - \\
& $Q^{(8)}_{qu}$ & -& - &  -  \\
& $Q^{(1)}_{qd}$ & -& - &  - \\
& $Q^{(8)}_{qd}$ & -& - & - \\
\bottomrule
\end{tabular}
\end{center}
\caption{All four-fermion operators that are invariant under CP and the $U(3)^5$ flavour symmetry. Ticks indicate that they contribute to the FCNC processes we consider. An asterisk $(*)$ signifies that the contribution is only indirect, via effects in input parameter measurements.}
\label{table:4fermion}
\end{table}

The SMEFT formalism expands upon the structure of the SM by allowing for additional, non-renormalizable operators. This introduces an additional perturbation series to the theory, expanding in inverse powers of the energy scale characterizing the new BSM physics, and an appropriate choice of expansion order must be made for both this new series as well as the usual series in gauge and Yukawa couplings already present in the SM. In this article we shall keep only the first non-trivial BSM contribution to the observables considered, which occurs at order $1/\Lambda^2$, where $\Lambda$ is again the new physics scale. 

We choose to work with the Warsaw basis~\cite{Grzadkowski:2010es} of dimension-6 operators for our calculations; this basis is particularly well suited to a loop-level calculation as higher-derivative operators have been systematically removed in its construction, and it is the only basis whose complete renormalization behaviour is known~\cite{Jenkins:2013zja,Jenkins:2013wua,Alonso:2013hga}. The full basis is given in the appendix in Table \ref{tab:basis}; we shall refer to operators by the names given in that table throughout the article, and denote the Wilson coefficient of operator $Q_a$ as $C_a$. 

We select operators by starting from a $U(3)^5$ flavour symmetry defined as
\begin{align}
U(3)_q \times U(3)_{u} \times U(3)_{d} \times U(3)_l \times U(3)_{e},
\end{align}
and under which the SM fermion fields have charges
\begin{align}
q&\sim (3,1,1,1,1), ~~ u \sim (1,3,1,1,1), ~~ d \sim (1,1,3,1,1),\\ l &\sim (1,1,1,3,1), ~~ e \sim (1,1,1,1,3).\nonumber
\end{align}
We are considering only the effects of flavour symmetric operators here, meaning operators which are overall singlets under the $U(3)^5$ symmetry. Many operators are straightforwardly forbidden from our analysis by this requirement; all the operators of classes 5 and 6 violate the flavour symmetry ansatz, as do the scalar-scalar interactions in class 8, and the operator $Q_{Hud}$. All operators which are invariant under $CP$ and the $U(3)^5$ flavour symmetry are listed in Tables~\ref{table:2fermion} and \ref{table:4fermion}, where we also indicate which down-type FCNC processes are affected by each operator at one loop.

 In the interest of compactness, we define the Wilson coefficients in the SMEFT to be dimensionful throughout, such that the dimension-six Lagrangian terms are written simply as
\begin{align}
\mathcal{L}_6=\sum_a C_a Q_a.
\end{align}
We drop flavour indices throughout our calculation, as our flavour symmetry assumption leads to the insistence that all the Wilson coefficient matrices in flavour space are identity-like, with the interesting exception of current-current four-fermion interactions of identical currents. Only three operators of that type contribute in our calculation: $Q_{ll}$  and $Q_{qq}^{(1)}$ contribute solely in the ``off-diagonal'' flavour combination which reads $\delta_{pt}\delta_{rs}$, and $Q_{qq}^{(3)}$ contributes in both allowed flavour combinations. The Wilson coefficients of the ``identity-like'' combination $\delta_{pr}\delta_{st}$ is denoted here unprimed ($C_{qq}^{(3)}$, $C_{qq}^{(1)}$, $C_{ll}$), while those of the ``off-diagonal'' combinations are primed ($C_{qq}^{(3)\prime}$, $C_{qq}^{(1)\prime}$, $C_{ll}^\prime$). 

In addition to restricting ourselves to the leading-order contributions in the new, EFT perturbation expansion, we shall also restrict our attention to contributions which arise at one-loop order (at most) in the SMEFT. Given the flavour assumptions we have made, there are no tree-level contributions to FCNC processes, with the exception of those arising from $Q_{qq}^{(1,3)}$. The fact that $Q_{qq}^{(1,3)}$ contain two quark currents make these the only operators that can mediate down-type quark flavour changing currents at tree level -- even if their Wilson coefficients are diagonal in the flavour basis -- due to the misalignment between the up- and down-type quark mass matrices characterised by the CKM. Upon rotating to the mass basis, therefore, interactions of the form $V_{ij}V^*_{kl}(\bar{u}^\alpha_i \gamma_\mu P_L d_j^\alpha)(\bar{u}_k^\beta \gamma_\mu P_L d_l^\beta)$ or $V_{ij}V^*_{kl} (\bar{u}^\alpha_i \gamma_\mu P_L d_j^\beta)(\bar{u}_k^\beta \gamma_\mu P_L d_l^\alpha)$ (where $\alpha$, $\beta$ are colour indices) are induced from these flavour singlet operators, similarly to the effect of integrating out the $W^\pm$ boson between two quark currents in the SM. In the vast majority of cases, though, the leading-order contribution of the SMEFT to FCNC processes starts at one loop under our assumptions. This is in contrast to the analysis of Ref.~\cite{Aebischer:2015fzz}, in which the SMEFT Wilson coefficients were considered in more generality, allowing most operators containing down-type quarks to contribute to $d_i \to d_j$ transitions at tree level. Hence Ref.~\cite{Aebischer:2015fzz} presents loop-level matching results only for operators containing a right-handed up type quark (some of which are excluded from our analysis since they are not $U(3)^5$ flavour singlets). These differing approaches ensure that many of the matching calculations presented here are new, but we compare with and refer to existing results in the literature wherever possible.

In the following we focus on $b\to s$ FCNC transitions for concreteness of notation, but (since the theory is flavour symmetric) our calculation applies equally well to $b\to d$ or $s\to d$ transitions as well, with the appropriate generation index replacements. Our goal in performing this calculation is to enable one-loop studies of the effects of flavour-symmetric SMEFT on down-type FCNC leptonic, semi-leptonic, or photonic decays and $\Delta F=2$ meson oscillations. 
We neglect loop-level matching to operators which only affect these processes via mixing, which leads to an additional suppression.

\section{Method and inputs}
\label{sec:methods}

Before embarking on the matching calculations, a choice must be made about which measurements to use to fix the free parameters of the theory. Measurements of inputs, for example the Fermi constant $G_F$, may be polluted by the effects of dimension-six operators in the SMEFT. Other dimension-six operators produce new contributions to the masses and mixings of gauge bosons and fermions when the Higgs takes its vev. Hence the coefficients of these operators will have knock-on effects wherever the inputs enter into other calculations. The input choice is especially important in the electroweak sector of the theory, where the presence of the operators $Q_{HD}$ and $Q_{HWB}$ breaks the usual SM relations between the Lagrangian parameters $v$, $g_1$, $g_2$, $\sin \theta$, $m_W$ and $m_Z$. These issues have been discussed at length in the literature (see e.g.~\cite{Burgess:1993vc,Han:2004az,Falkowski:2014tna,Berthier:2015oma,Brivio:2017bnu}). 

In the main text of this paper, we present our results in a scheme in which the set of input measurements are
\begin{equation}
\left\lbrace \hat{m}_W, \hat{m}_Z, \hat{G}_F, \hat{m}_t, \hat{m}_b, \hat{\alpha}_s, \hat{V}_{CKM}  \right\rbrace.
\end{equation}
We denote these measured inputs, as well as parameters derived from them via SM relations, with a hat \cite{Brivio:2017bnu}:
\begin{align}
\hat{v}&=\frac{1}{2^{1/4}\sqrt{\hat{G}_F}}, ~~~\hat{g}_1=2\cdot 2^{1/4} \hat{m}_Z\sqrt{\hat{G}_F\left( 1-\frac{\hat{m}_W^2}{\hat{m}_Z^2}\right)},~~~\hat{g}_2=2\cdot 2^{1/4} \hat{m}_W\sqrt{\hat{G}_F},\nonumber\\
\hat{y}_t &= \frac{\sqrt{2}\hat{m}_t}{\hat{v}},~~~\hat{y}_b = \frac{\sqrt{2}\hat{m}_b}{\hat{v}}, ~~~\hat{g}_s = \sqrt{4\pi \hat{\alpha}_s}.
\end{align}
Within our $U(3)^5$ flavour assumption, the mapping from the measured inputs $\hat{\alpha}_s$ and $\hat{V}_{CKM}$ to Lagrangian parameters goes through similarly to in the SM.\footnote{See Ref.~\cite{Descotes-Genon:2018foz} for a more general treatment of the CKM matrix within the SMEFT.} However, things are slightly less trivial for the electroweak sector. In this case there are three free Lagrangian parameters, which we take to be the gauge couplings $\bar{g}_1$ and $\bar{g}_2$, and the electroweak vev $\bar{v}$, where the bars indicate that these are SMEFT Lagrangian parameters. Once these are fixed by solving for the input measurements $\hat{G}_F$, $\hat{m}_W$ and $\hat{m}_Z$, they can be written as the sum of the respective hatted parameters and a shift which depends on SMEFT dimension-six Wilson coefficients:
\begin{align}
\bar{g}_1 = \hat{g}_1 + \delta g_1,\nonumber \\
\bar{g}_2 = \hat{g}_2 + \delta g_2,\label{eqn:shiftsdefs} \\
\bar{v} = \hat{v} + \delta v.\nonumber
\end{align}
For our choice of inputs, the operators that enter into the shifts $\delta g_1$, $\delta g_2$ and $\delta v$ are $C_{HWB}$, $C_{HD}$, $C_{Hl}^{(3)}$ and $C_{ll}^\prime$. Then the procedure for deriving the Feynman rules can be understood via the following steps:
\begin{enumerate}
\item Write the Lagrangian in terms of canonically normalised mass eigenstates, and the three free electroweak parameters ($\bar{g}_1$, $\bar{g}_2$, $\bar{v}$) 
\item Derive Feynman rules in terms of these three parameters
\item Write the free parameters in terms of measured inputs and the dimension-six shifts ($\delta g_1$, $\delta g_2$ and $\delta v$) and substitute them into Feynman rules, consistently retaining terms of order $1/\Lambda^2$
\end{enumerate}
Steps 1 and 2 have been done in Ref.~\cite{Dedes:2017zog}, and step 3 can be trivially performed from the Feynman rules in that reference using Eqns.~\eqref{eqn:shiftsdefs},
and remembering that the gauge boson masses in the SMEFT are\footnote{For derivations of these see e.g.~Ref.~\cite{Alonso:2013hga}}
\begin{align}
m_W^2 &= \frac{\bar{g}_2^2 \bar{v}^2}{4},\\
m_Z^2 &= \frac{\bar{v}^2}{4}\left(\bar{g}_1^2+\bar{g}_2^2 \right)+\frac{1}{8}\bar{v}^4C_{HD}\left(\bar{g}_1^2+\bar{g}_2^2 \right)+\frac{1}{2}\bar{v}^4 \bar{g}_1 \bar{g}_2 C_{HWB}.
\end{align}
Since the measured quark masses $\hat{m}_t$ and $\hat{m}_b$ are taken as inputs, the Yukawa couplings $\bar{y}_t$ and $\bar{y}_b$ are affected by the shift $\delta v$ as
\begin{align}
\bar{y}_t &= \sqrt{2}\frac{\hat{m}_t}{\bar{v}} = \hat{y}_t\left(1- \delta v\right),\\
\bar{y}_b &= \sqrt{2}\frac{\hat{m}_b}{\bar{v}} = \hat{y}_b\left(1- \delta v\right).
\end{align}
All other fermion masses are set to zero in our calculation.\footnote{with the exception of including leading charm mass effects in meson mixing coefficients, for application to kaon mixing} We provide more explicit details of the procedure -- including the expressions for $\delta g_1$, $\delta g_2$ and $\delta v$, as well as quoting our results in a different scheme in which $\lbrace \hat \alpha_{em}, \hat m_Z, \hat G_F\rbrace$ are the electroweak inputs -- in App.~\ref{sec:inputsappendix}.

The WET effective Hamiltonian for $\Delta B = \Delta S = 1$ transitions to which the symmetric SMEFT matches is identical to the WET basis of the SM. This is a consequence of our flavour symmetry assumption. 
\begin{equation}
\mathcal{H}_{\text{eff}}^{|\Delta B|=|\Delta S|=1}= \frac{4\hat{G}_F}{\sqrt{2}}\left[ - \frac{1}{(4\pi)^2}\hat{V}_{ts}^* \hat{V}_{tb}\sum_{i=3}^{10}C_i \mathcal{O}_i +\sum_{q=u,c}\hat{V}_{qs}^* \hat{V}_{qb}\, ( C_1 \mathcal{O}^q_1 + C_2 \mathcal{O}^q_2 ) \right],
\end{equation}
with

\begin{align}
\mathcal{O}^q_1&= (\bar b^\alpha \gamma_\mu P_L q^\beta)(\bar q^\beta \gamma^\mu P_L s^\alpha), \nonumber\\
\mathcal{O}^q_2 &=(\bar b^\alpha \gamma_\mu P_L q^\alpha)(\bar q^\beta \gamma^\mu P_L s^\beta), \nonumber\\
\mathcal{O}_7 &=\hat{e} \hat{m}_b\left(\bar{s}\sigma^{\mu\nu}P_R b \right)F_{\mu\nu},\nonumber\\
\mathcal{O}_8 &=\hat{g}_s\hat{m}_b\left(\bar{s}\sigma^{\mu\nu}T^AP_R b \right)G_{\mu\nu}^A,\nonumber\\
\mathcal{O}_9 &=\hat{e}^2\left( \bar{s}\gamma^{\mu}P_L b\right) \left( \bar{\ell}\gamma_{\mu} \ell \right), \nonumber\\
\mathcal{O}_{10} &=\hat{e}^2 \left( \bar{s}\gamma^{\mu}P_L b\right) \left( \bar{\ell}\gamma_{\mu}\gamma_5 \ell \right). 
\end{align}
where $\alpha$, $\beta$ are colour indices. For the definition of the QCD penguin operators $\mathcal{O}_{3,4,5,6}$ we refer to Ref.~\cite{Buchalla:1995vs}.
In the SM, $C_2$ receives tree level contributions, while $C_{3-10}$ are generated only at loop level. We will find that our flavour symmetry assumption ensures that a similar matching pattern arises in the SMEFT, although we additionally get a tree level contribution to $C_1$.  As discussed before, we neglect loop-level matching to four-quark operators, since their effects in these processes is only via mixing and, thus, are suppressed compared to the direct one-loop matching contributions to $C_7 - C_{10}$.

The WET effective Hamiltonian for $\Delta B = \Delta S = 2$ transitions is again identical to the WET basis of the SM,
\begin{align}
\mathcal{H}_{\text{eff}}^{|\Delta B|=|\Delta S|=2}&=\frac{\hat{G}_F^2 \hat{m}_W^2}{16\pi^2} \left(\bar{s}^\alpha_L \gamma^\mu b^\alpha_L \right)(\bar{s}^\beta_L \gamma^\mu b^\beta_L)\nonumber \\
&\times \left(\lambda_t^2\, C^s_{1,mix}(x_t)+\lambda_c^2 \,C^s_{1,mix}(x_c) +2\, \lambda_c \lambda_t\, C^s_{1,mix}(x_t, x_c)\right), 
\end{align}
where $\alpha$ and $\beta$ are colour indices, and $\lambda_i=\hat{V}_{is}^* \hat{V}_{ib}$. The coefficients $C^s_{1,mix}$ are functions of $x_i = m_i^2/m_W^2$, and only the first term $\lambda_t^2\, C^s_{1,mix}(x_t)$ is non-negligible in the case of $B_{s}$ (and $B_{d}$) mixing. However we include the functions $C^s_{1,mix}(x_c)$ and $C^s_{1,mix}(x_t,x_c)$ here -- and quote their values (to linear order in $x_c \ll 1$) in the main text -- to allow application of our matching results via trivial flavour index replacements to kaon mixing, where these terms are important.

\section{Results}
\label{sec:results}
In this section we present our results for the matching of the $U(3)^5$ flavour and CP symmetric SMEFT theory onto the coefficients of the WET. All WET Wilson coefficients are at the electroweak scale $m_W$. We define Wilson coefficients in the SMEFT at the arbitrary scale $\mu$ and do not resum the logarithmic divergences of form $\log\left(\frac{\mu}{m_W}\right)$, leaving them explicit in our calculation for comparison with the anomalous dimension matrix of \cite{Jenkins:2013zja,Jenkins:2013wua,Alonso:2013hga}, with which we find agreement. We separate our results by SMEFT operator, or groups of similar operators, and we only present non-zero results. Our calculations have been done in $R_\xi$ gauge using dimensional regularisation and we use the $\overline{MS}$ prescription to remove divergences. In all cases 
we have confirmed that the separate contributions calculated here are gauge parameter independent.
Where possible, we compare our results to those obtained previously in the literature. In all diagrams, orange blobs represent insertions of SMEFT operators, and unlabelled internal fermion lines are $u/c/t$ quarks.

To first order in $1/\Lambda^2$, the barred and hatted parameters (e.g.~$\bar{g}_1$, $\hat{g}_1$ as introduced in Sec.~\ref{sec:methods}) are equal when they are multiplied by a SMEFT Wilson coefficient, so in the following we simply drop the hats and bars for simplicity. However we emphasise that the results presented here include the effects of input parameter shifts, and we are taking $\left\lbrace m_W, m_Z, G_F \right\rbrace$ as the set of electroweak input parameters, as explained in Sec.~\ref{sec:methods}.

\subsection{\boldmath{$Q_{qq}^{(1)}$ and $Q_{qq}^{(3)}$}}
\label{sec:Qqq}
\begin{figure}
\begin{center}
\includegraphics[height=1.8cm]{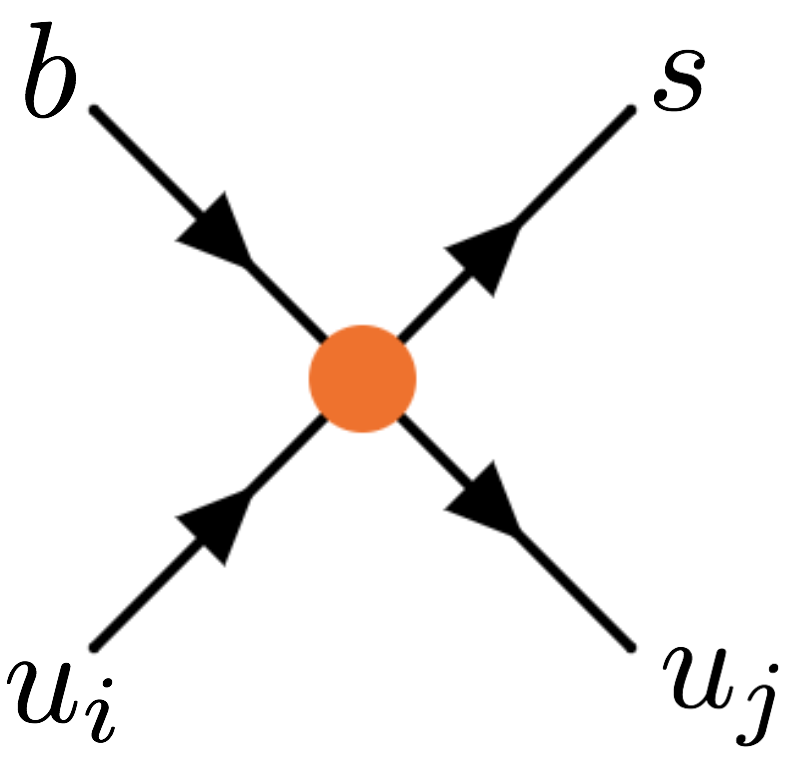}~~~~
\includegraphics[height=1.8cm]{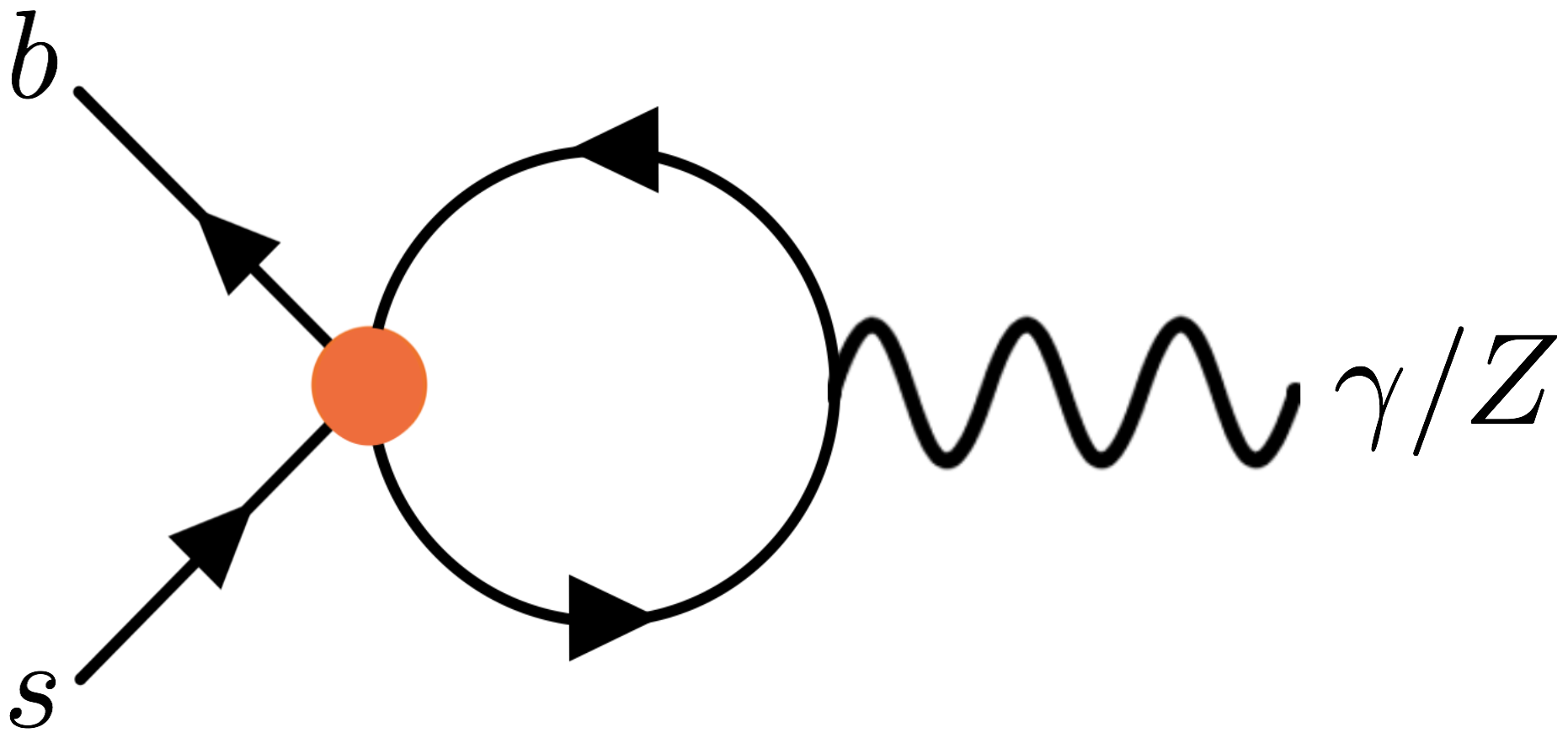}
\caption{Diagrams generating contributions to the 4-quark WET coefficients $C_{1,2}$ (left) and to $b\to s l^+ l^-$ (right) from the SMEFT operators $Q_{qq}^{(1)}$ and $Q_{qq}^{(3)}$. \label{Cqq}}
\end{center}
\end{figure}
\begin{figure}
\begin{center}
\includegraphics[height=1.8cm]{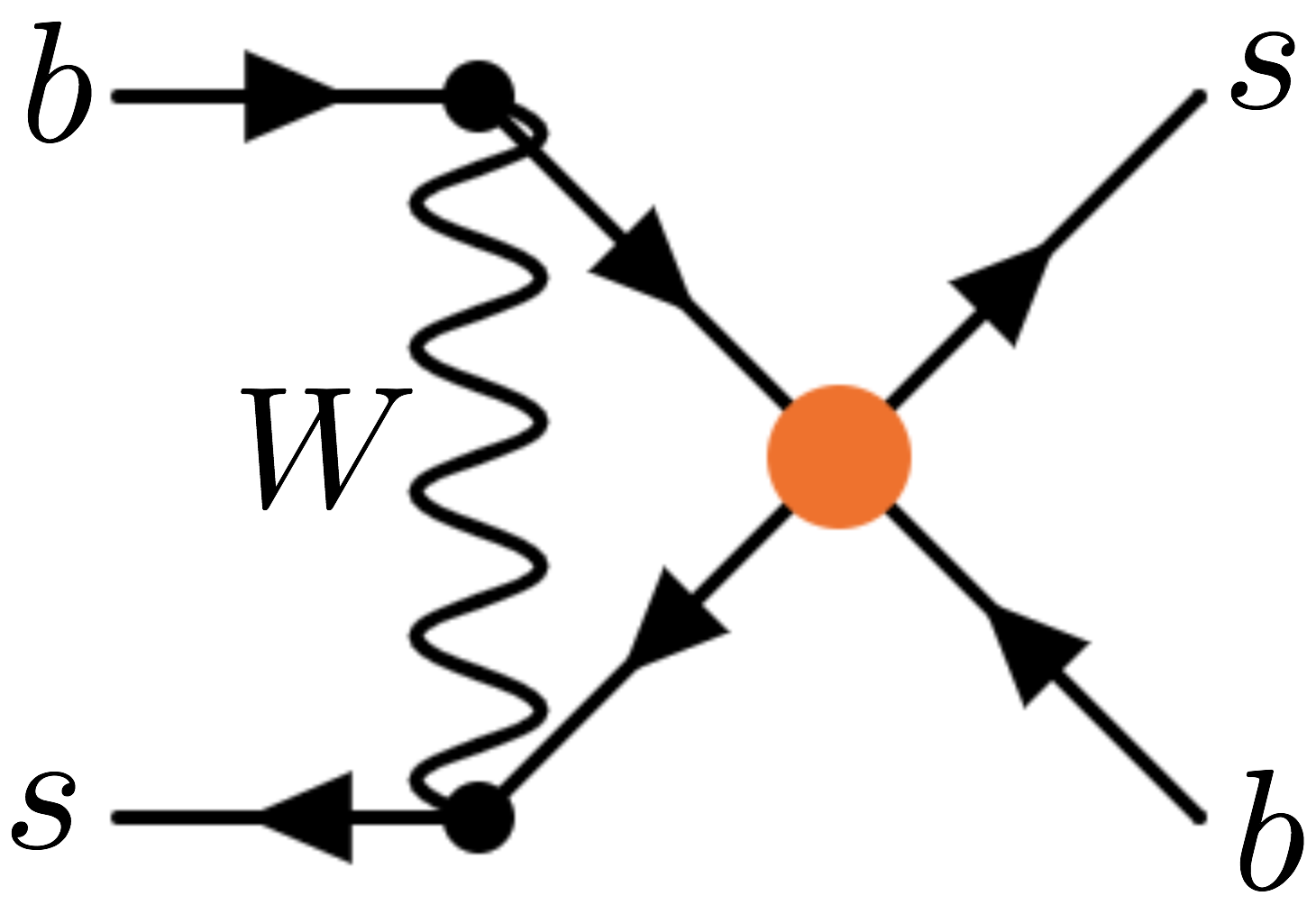}~~~~
\includegraphics[height=1.8cm]{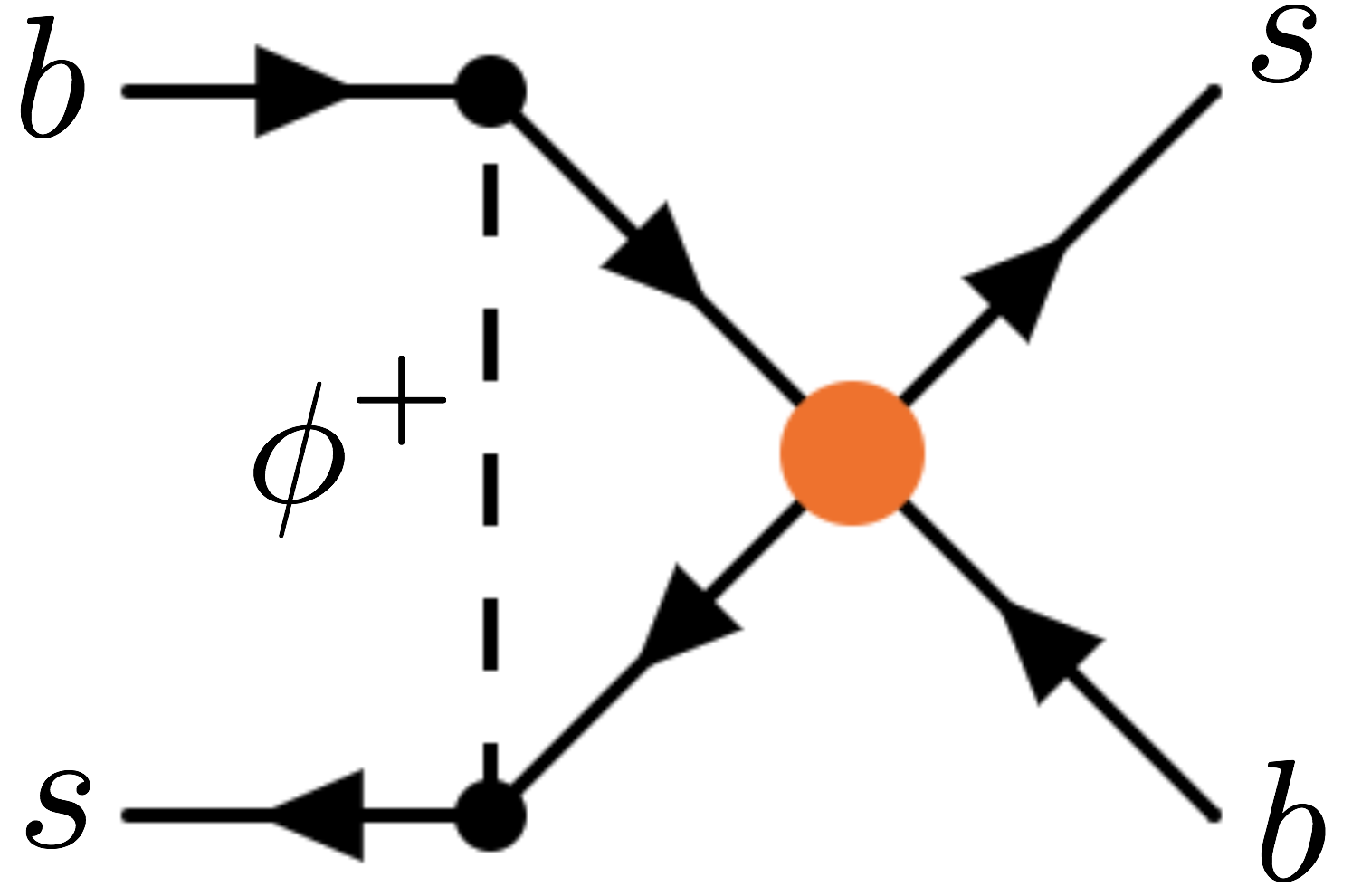}
\caption{Diagrams generating contributions to $B_s$ mixing from the SMEFT operators $Q_{qq}^{(1)}$ and $Q_{qq}^{(3)}$. \label{Cqqmix}}
\end{center}
\end{figure}
These are the only operators in our theory which generate a contribution to the $b\to s$ transition at tree level, shown in Fig.~\ref{Cqq}. As mentioned in Sec.~\ref{sec:flavoursymm}, there are two ways of contracting the quark doublets to make a flavour singlet in both operators, so we can define four independent Wilson coefficients, $C_{qq}^{(3)}$, $C_{qq}^{(3)\prime}$, $C_{qq}^{(1)}$ and $C_{qq}^{(1)\prime}$, within our $U(3)^5$ invariant theory.

The contribution to the coefficients of the 4-quark WET operators $\mathcal{O}_1$ and $\mathcal{O}_2$ is
\begin{align}
C_1 &=v^2 (C_{qq}^{(3)\prime}-C_{qq}^{(1)\prime}),\\
C_2 &=-2 v^2 C_{qq}^{(3)}.
\end{align}
Contributions to $C_9$ and $C_{10}$ are generated by the second diagram of Fig.~\ref{Cqq}. We find
\begin{align}
C_9&= -v^2C_{qq}^{(3)}\left(\frac{8}{9}+\frac{x_t}{2}\frac{1-4 s_\theta^2}{s_\theta^2} \right)\left(1+\log \frac{m_t^2}{\mu^2}  \right)\nonumber\\&+v^2N_c (C_{qq}^{(3)\prime}-C_{qq}^{(1)\prime})\left(\frac{4}{9}+\frac{x_t}{4}\frac{1-4 s_\theta^2 }{s_\theta^2} \right)\log \frac{m_t^2}{\mu^2}, \\
C_{10}&= \frac{1}{2}\frac{1}{s_\theta^2} v^2C_{qq}^{(3)}x_t\left(1+\log \frac{m_t^2}{\mu^2}  \right)-\frac{N_c}{4}\frac{1}{s_\theta^2} v^2 (C_{qq}^{(3)\prime}-C_{qq}^{(1)\prime})x_t\log \frac{m_t^2}{\mu^2}.
\end{align}
where $N_c=3$ is the number of QCD colours.
These operators also generate contributions to $B_s$ mixing from the diagrams in Fig.~\ref{Cqqmix}. These give
\begin{align}
C_{1,\text{mix}}^s (x_t)&=-v^2\left(2C_{qq}^{(3)} +(C_{qq}^{(1)\prime}-C_{qq}^{(3)\prime})\right)\,x_t,\\
C_{1,\text{mix}}^s (x_c)&=-v^2\left(2C_{qq}^{(3)} +(C_{qq}^{(1)\prime}-C_{qq}^{(3)\prime})\right)\,x_c,\\
C_{1,\text{mix}}^s (x_t,x_c)&=v^2\left(2C_{qq}^{(3)} +(C_{qq}^{(1)\prime}-C_{qq}^{(3)\prime})\right)\, x_c \log \frac{x_c}{x_t}.
\end{align}

\subsection{\boldmath{$Q_{lq}^{(1)}$, $Q_{lq}^{(3)}$, $Q_{eu}$, $Q_{lu}$ and $Q_{qe}$}}
\begin{figure}
\begin{center}
\includegraphics[height=1.8cm]{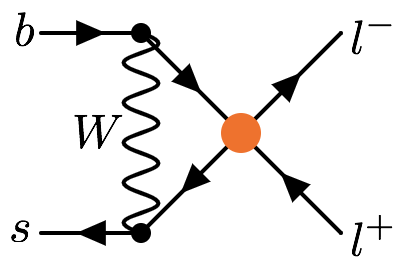}~~
\includegraphics[height=1.8cm]{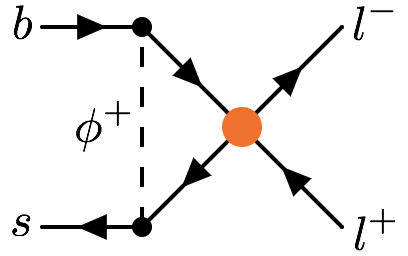}~~
\includegraphics[height=2cm]{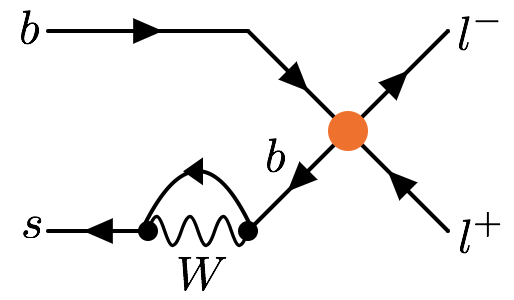}~~
\includegraphics[height=2cm]{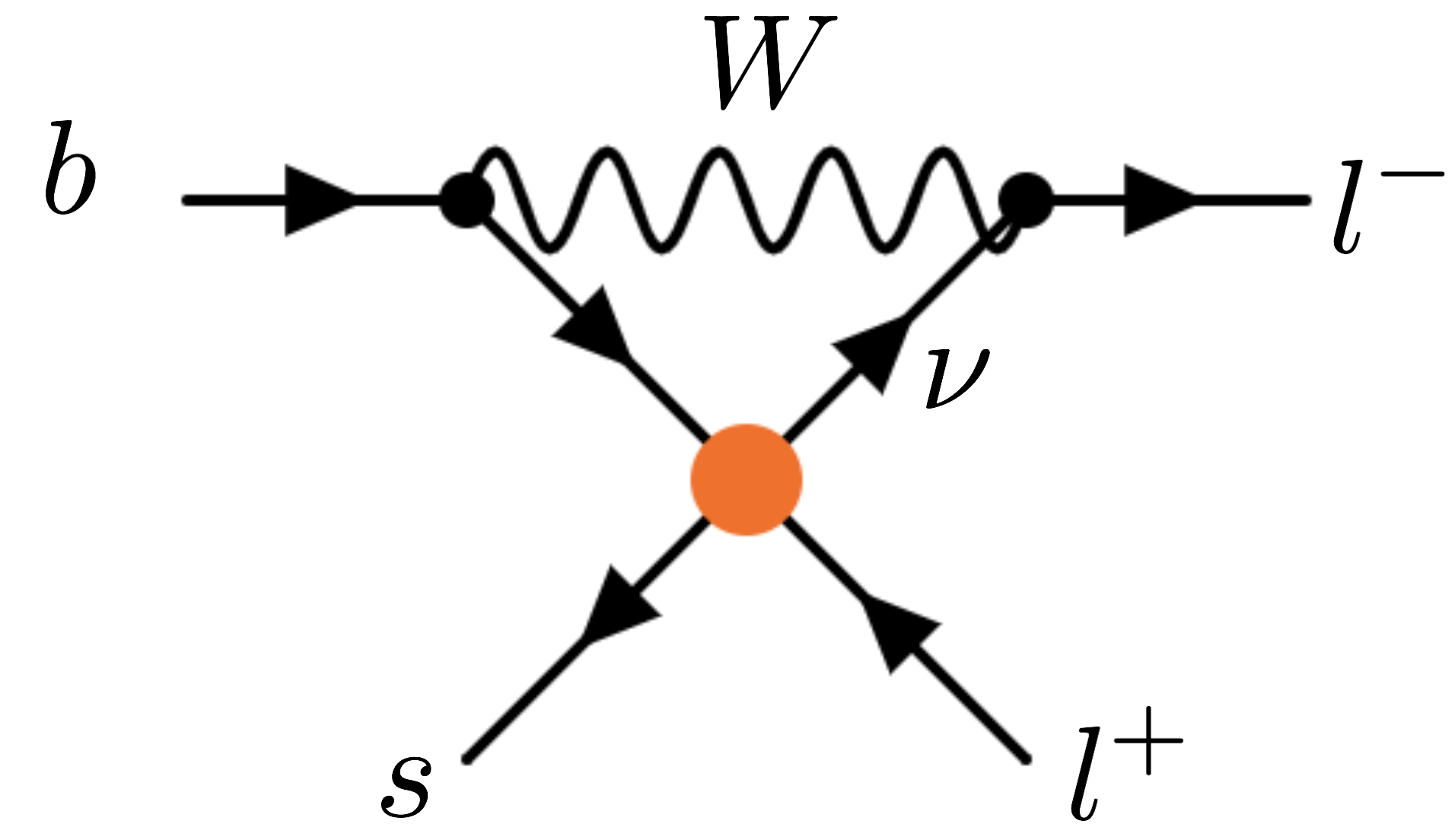}
\caption{Diagrams generating contributions to $b\to s l^+ l^-$ from $Q_{eu}$ and $Q_{lu}$ (first two diagrams only),  $Q_{lq}^{(1)}$ and $Q_{qe}$ (first three diagrams and similar) and $Q_{lq}^{(3)}$ (all four diagrams and similar). \label{Clq}}
\end{center}
\end{figure}

These four-fermion operators contribute to $b\to s l^+ l^- $ processes via the diagrams shown in Fig.~\ref{Clq}. The contributions are
\begin{align}
C_{9}&=\frac{v^2}{s_{\theta}^2}(C_{eu}+C_{lu}-C_{lq}^{(1)}-C_{qe})I(x_t)-\frac{v^2}{s_{\theta}^2}C_{lq}^{(3)} I^{lq}(x_t),\\
C_{10}&=\frac{v^2}{s_{\theta}^2}(C_{eu}-C_{lu}+C_{lq}^{(1)}-C_{qe})I(x_t)+\frac{v^2}{s_{\theta}^2}C_{lq}^{(3)} I^{lq}(x_t),
\end{align}
where 
\begin{align}
I(x_t) &=\frac{x_t}{16}\left[ -\log \frac{m_W^2}{\mu^2}+\frac{x_t-7}{2(1-x_t)}-\frac{x_t^2-2x_t+4}{(1-x_t)^2}\log x_t\right], \label{eqn:I}\\
I^{lq}(x_t)&= \frac{x_t}{16}\left[ -\log \frac{m_W^2}{\mu^2}+\frac{1-7x_t}{2(1-x_t)}-\frac{x_t^2-2x_t+4}{(1-x_t)^2}\log x_t\right]. \label{eqn:Ilq}
\end{align}
Our results for $C_{eu}$ and $C_{lu}$ are in agreement with Ref.~\cite{Aebischer:2015fzz}.

\subsection{\boldmath{$Q_{Hl}^{(1)}$ and $Q_{He}$}}

\begin{figure}
\begin{center}
\includegraphics[height=2cm]{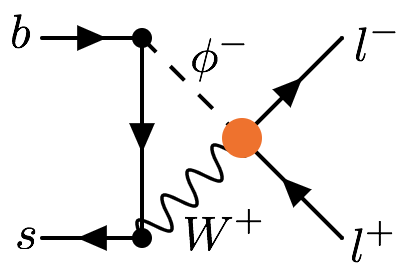}~~
\includegraphics[height=2cm]{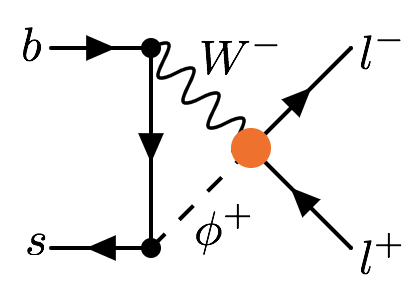}~~
\includegraphics[height=2cm]{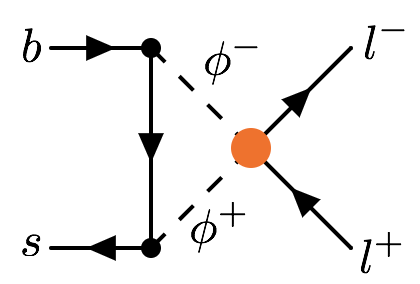}~~
\includegraphics[height=1.8cm]{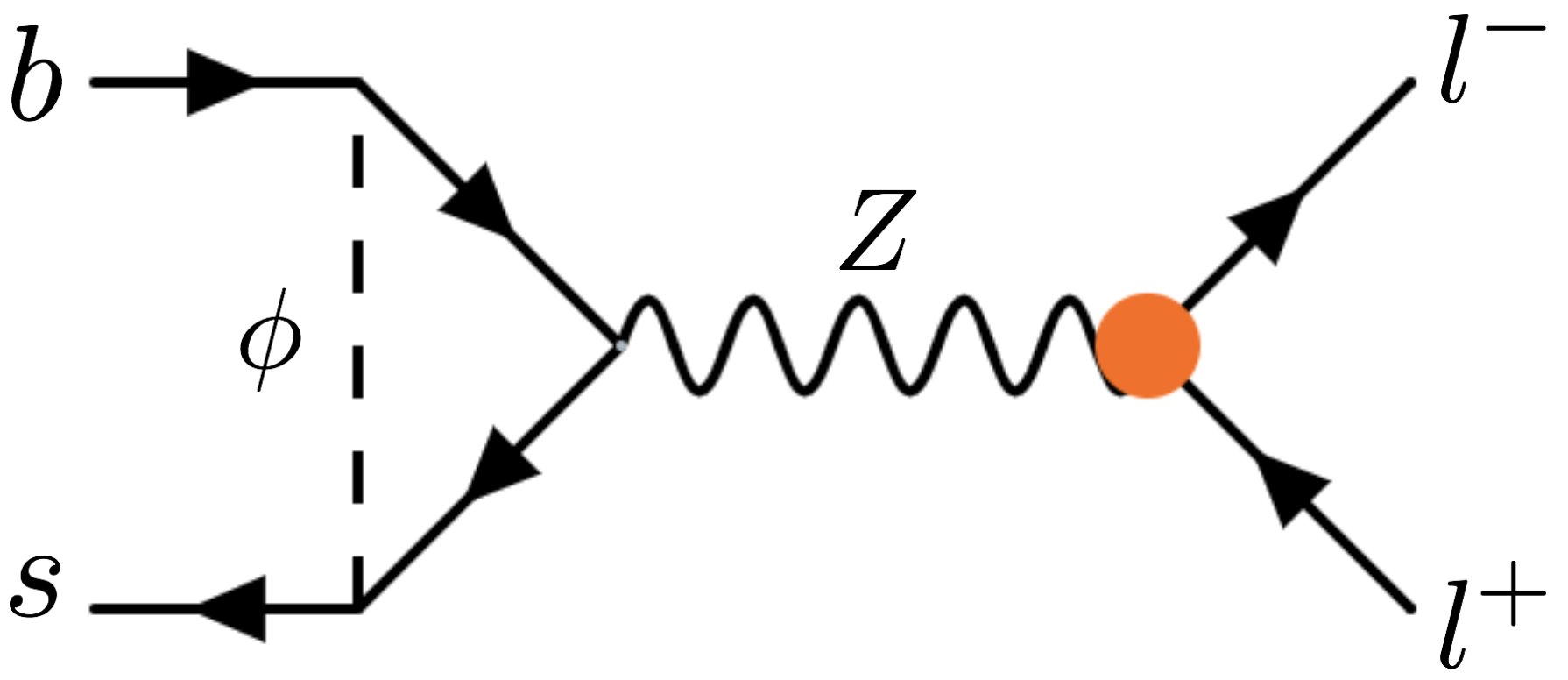}
\caption{Diagrams generating contributions to $b\to s l^+ l^-$ from $Q_{Hl}^{(1)}$ and $Q_{He}$ operators. The fourth diagram should be taken to include all other $Z$ penguin diagrams (including those with self-energies on external legs) where these operators affect the $Zl^+ l^-$ vertex. \label{CHe}}
\end{center}
\end{figure}
These operators produce effects in $b\to s l^+ l^-$ via the diagrams shown in Fig.~\ref{CHe}, giving
\begin{align}
C_{9}&= -\frac{v^2}{s_{\theta}^2}\left(C_{Hl}^{(1)} +C_{He} \right)I(x_t),\\
C_{10}&= \frac{v^2}{s_{\theta}^2}\left(C_{Hl}^{(1)} -C_{He} \right)I(x_t),
\end{align}
where $I(x_t)$ is defined in Eqn.~\eqref{eqn:I}.

\subsection{\boldmath{$Q_{Hq}^{(1)}$ and $Q_{Hu}$}}

\begin{figure}
\begin{center}
\includegraphics[height=1.7cm]{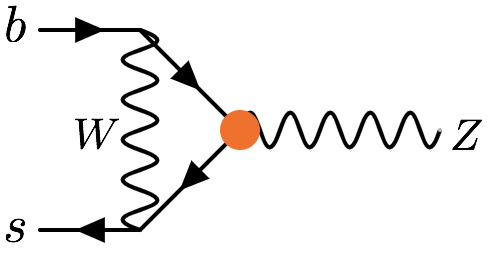}~~~~
\includegraphics[height=1.7cm]{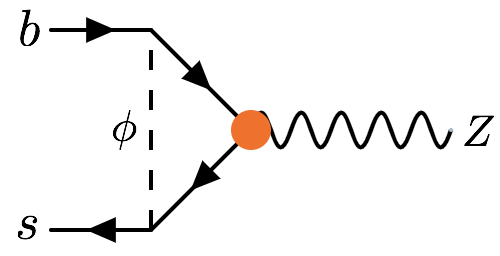}
\caption{Diagrams generating contributions to $b\to s l^+ l^-$ from $Q_{Hq}^{(1)}$ and $Q_{Hu}$ operators.\label{CHu}}
\end{center}
\end{figure}
These operators effectively just change the $Z\bar{u}_iu_i$ coupling and hence only enter in the $Z$ penguin diagrams shown in Fig.~\ref{CHu}. The contributions are
\begin{align}
C_9&= v^2\frac{(1-4s_{\theta}^2)}{s_{\theta}^2}\left(C_{Hu}-C_{Hq}^{(1)} \right) I(x_t),\\
C_{10}&=\frac{v^2}{s_{\theta}^2}\left(C_{Hu}-C_{Hq}^{(1)} \right) I(x_t),
\end{align}
where $I(x_t)$ is defined in Eqn.~\eqref{eqn:I}. The $C_{Hu}$ result is in agreement with Ref.~\citep{Aebischer:2015fzz}.

\subsection{\boldmath{$Q_{Hq}^{(3)}$}}
\begin{figure}
\begin{center}
\includegraphics[height=1.7cm]{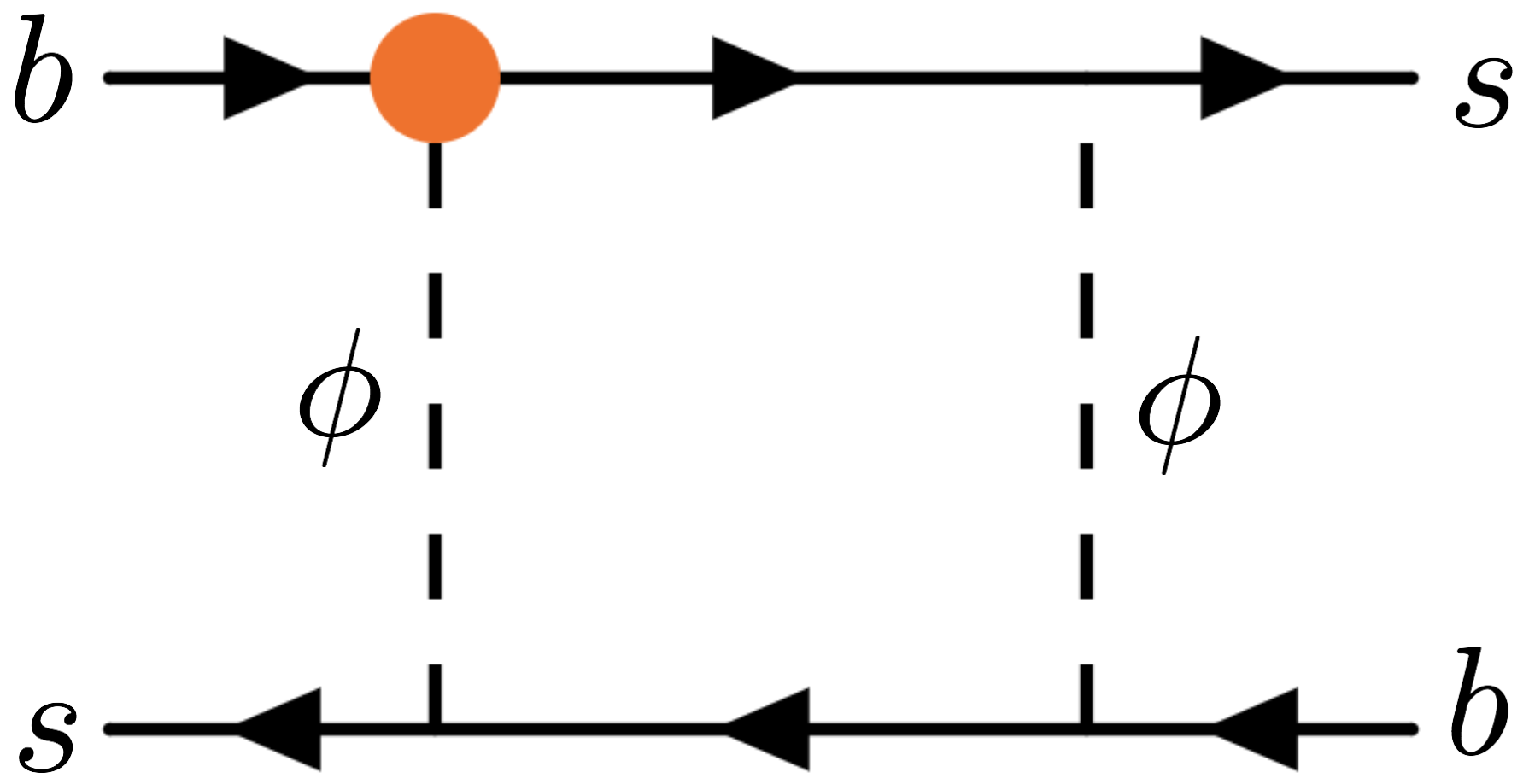}~~~~
\includegraphics[height=1.7cm]{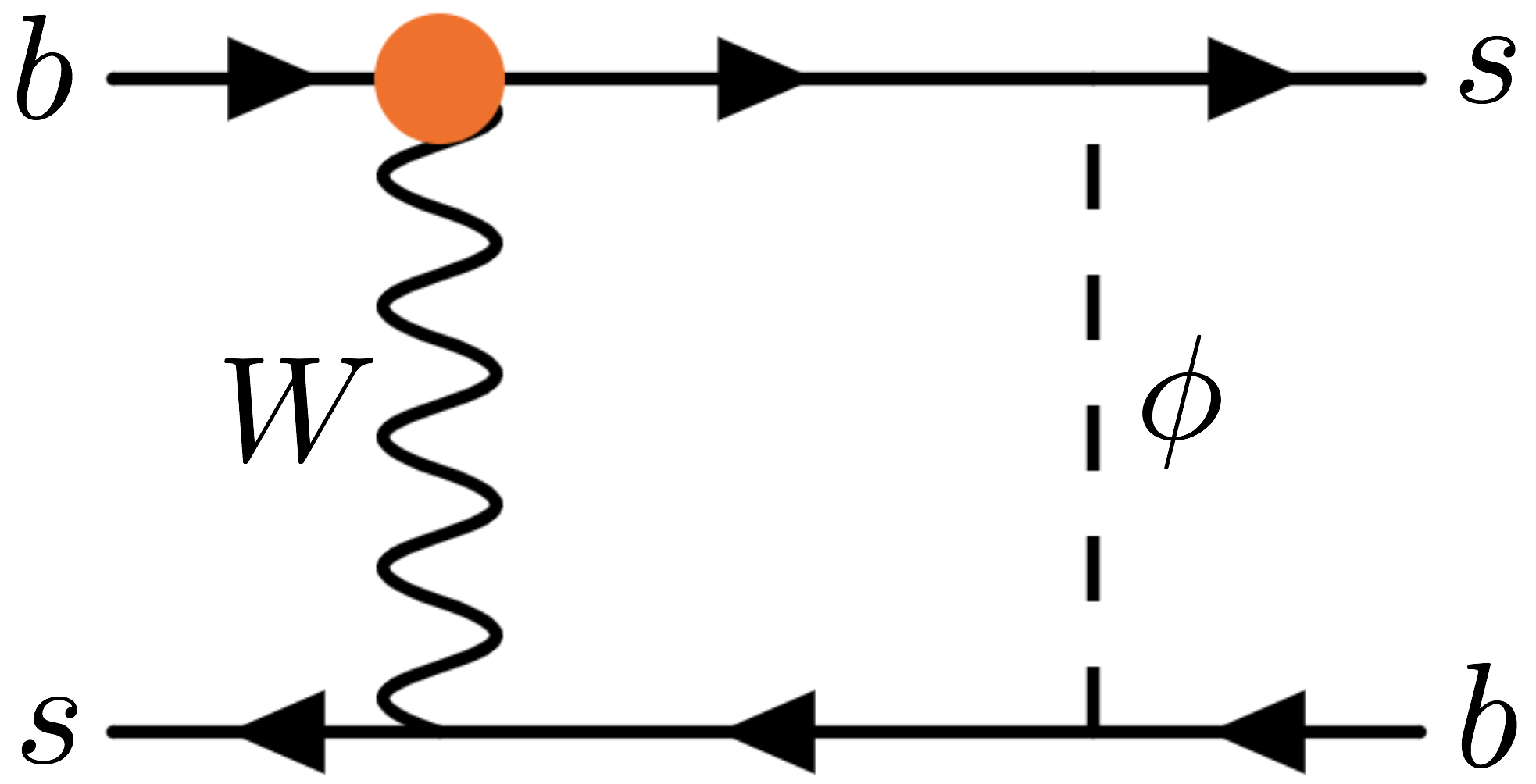}~~~~
\includegraphics[height=1.7cm]{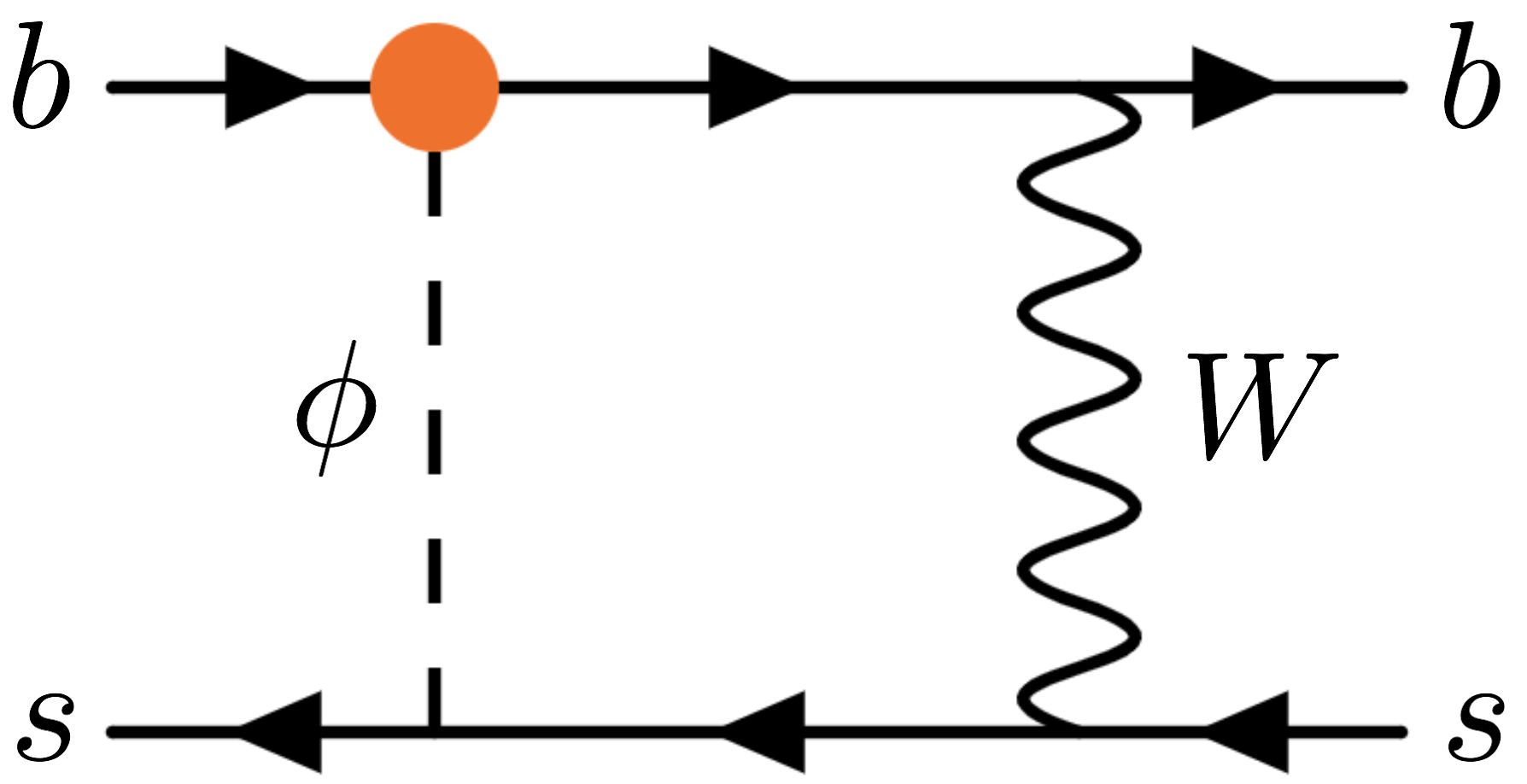}
\caption{Diagrams generating contributions to $B_s$ mixing from the $Q_{Hq}^{(3)}$ operator. Diagrams related to these by symmetry, as well as diagrams in which the operator connects to an $s$ quark leg rather than a $b$ quark leg, should be taken to be included. \label{CHq3mix}}
\end{center}
\end{figure}
\begin{figure}
\begin{center}
\includegraphics[height=1.5cm]{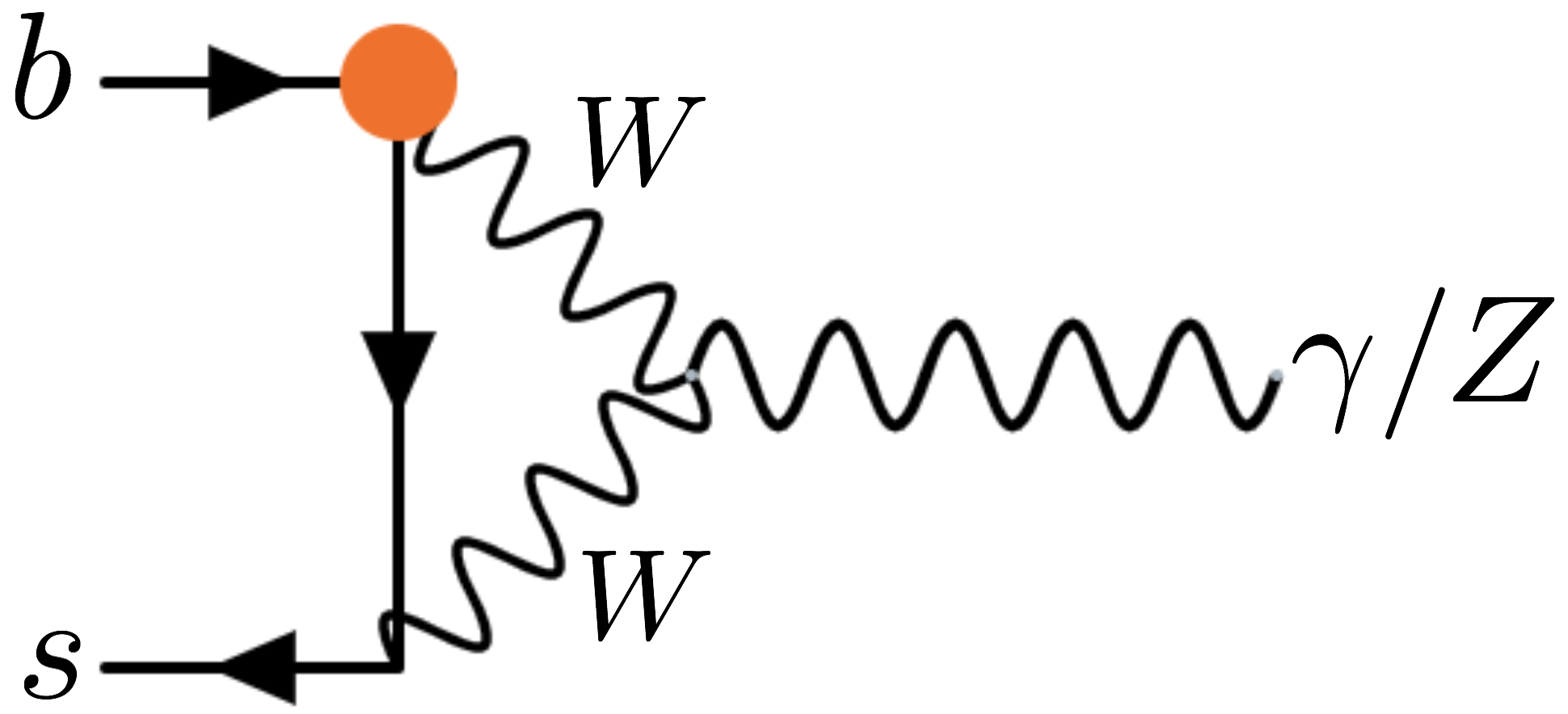}~~
\includegraphics[height=1.5cm]{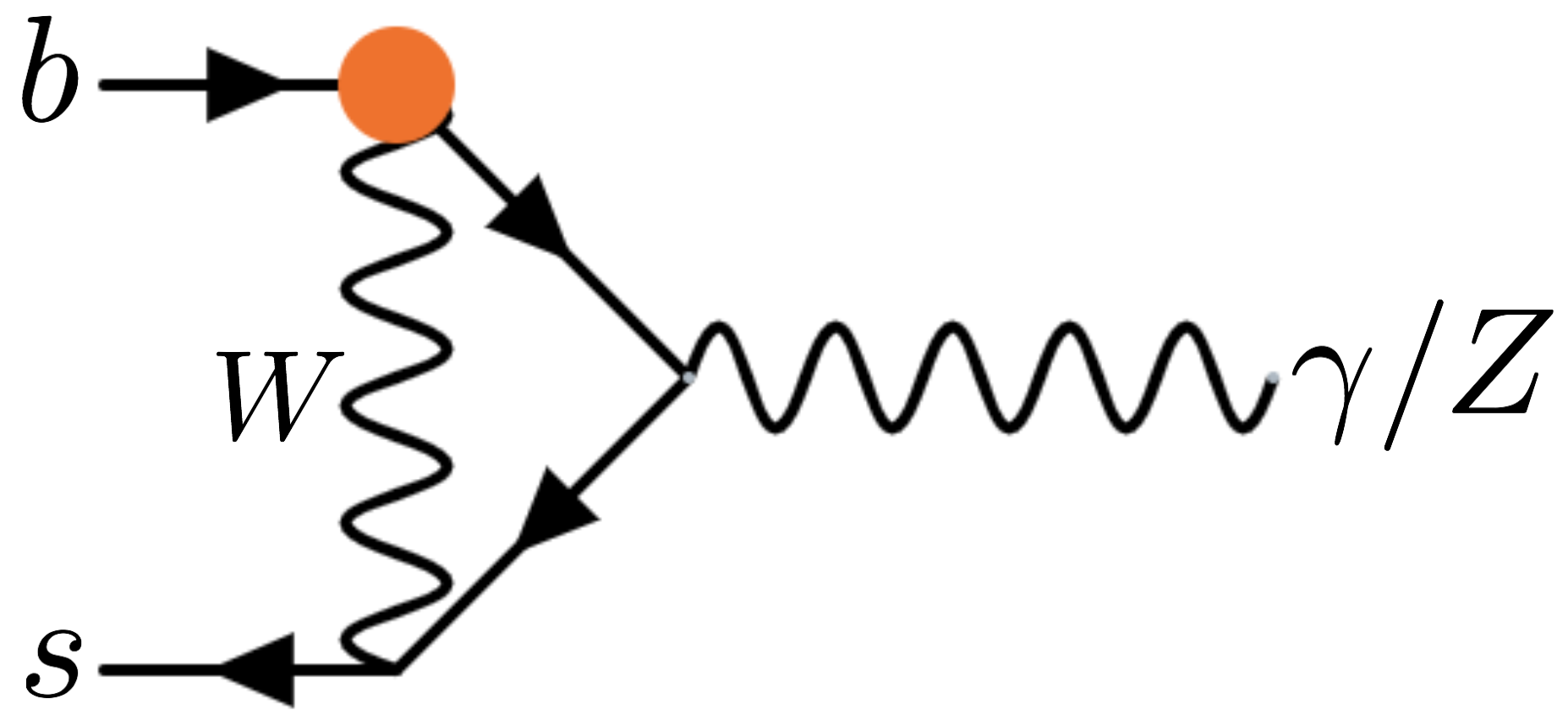}~~
\includegraphics[height=1.5cm]{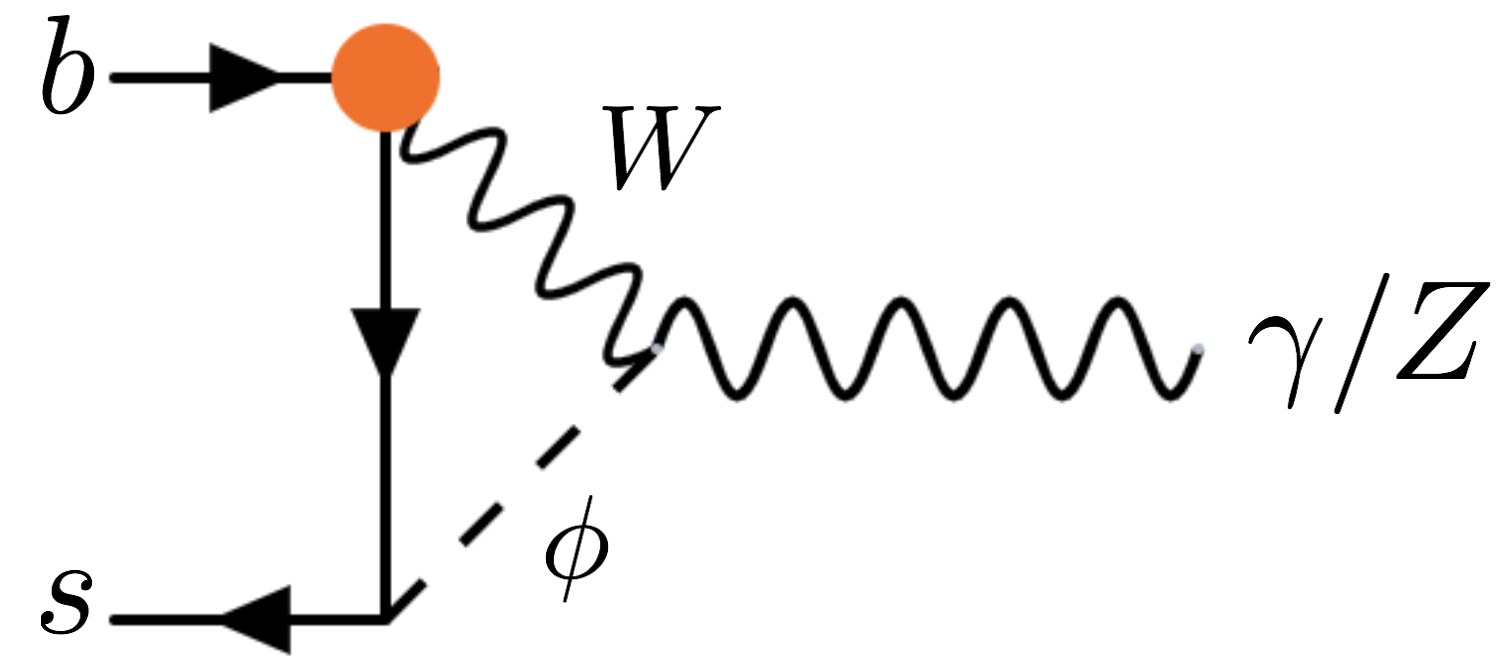}~~
\includegraphics[height=1.5cm]{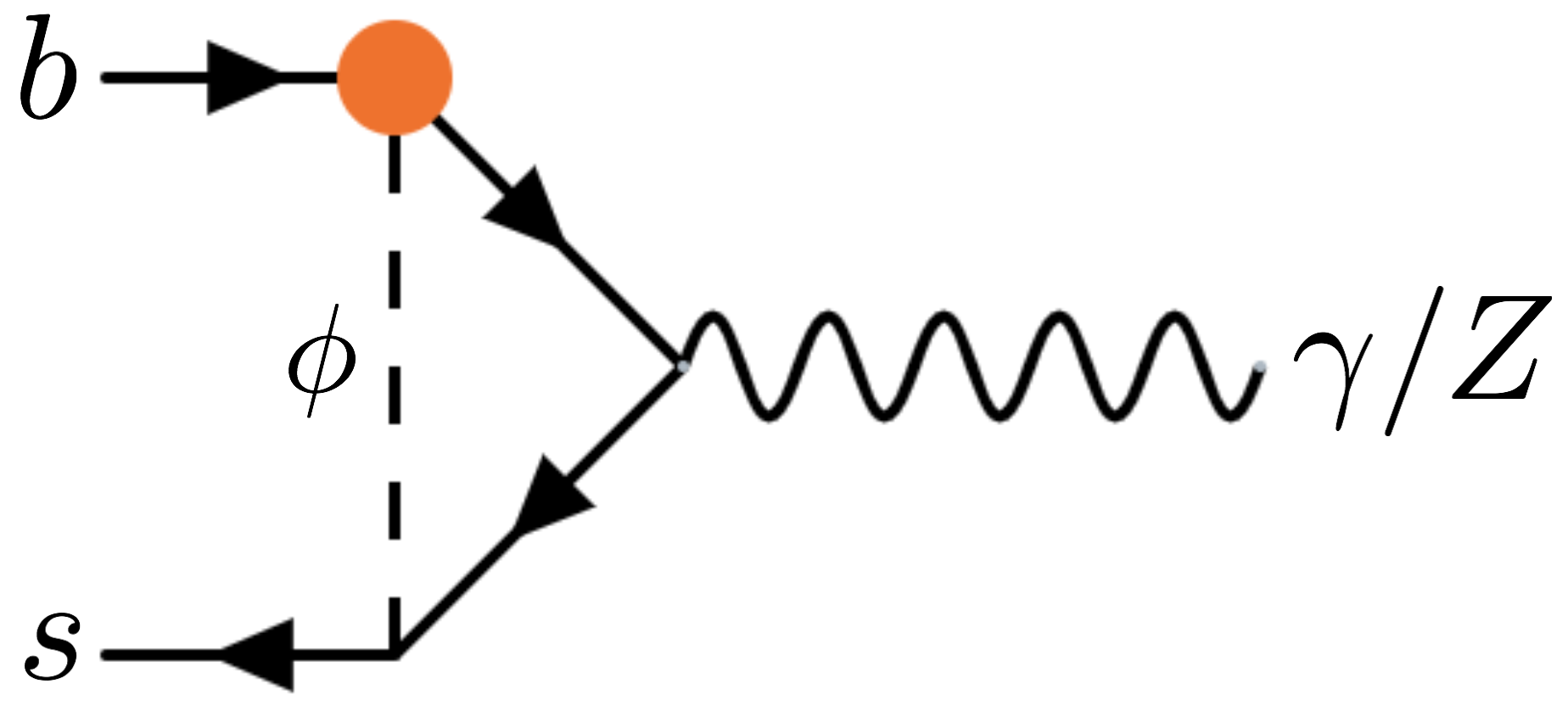}\\
\vspace{0.8cm}
\includegraphics[height=1.6cm]{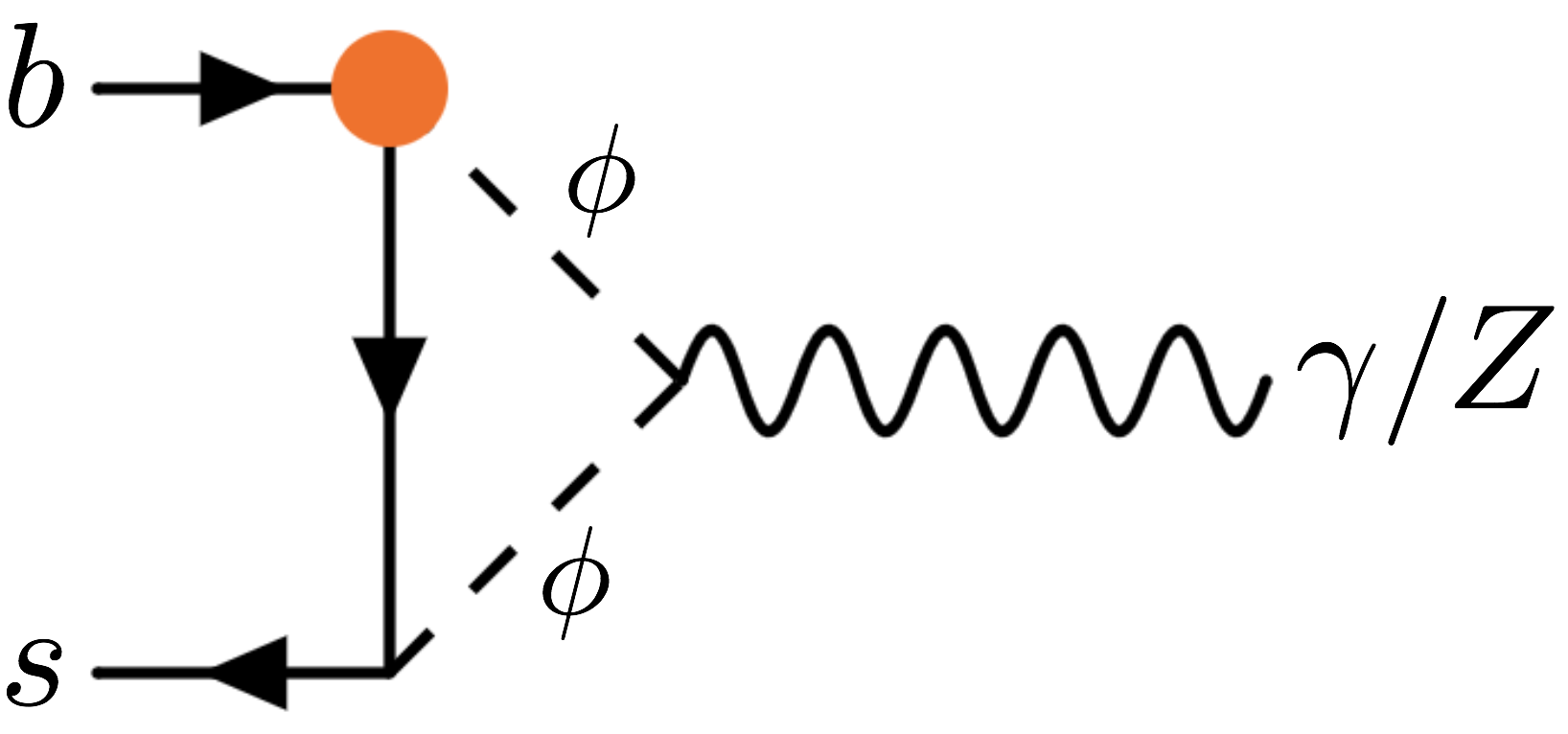}~~
\includegraphics[height=1.5cm]{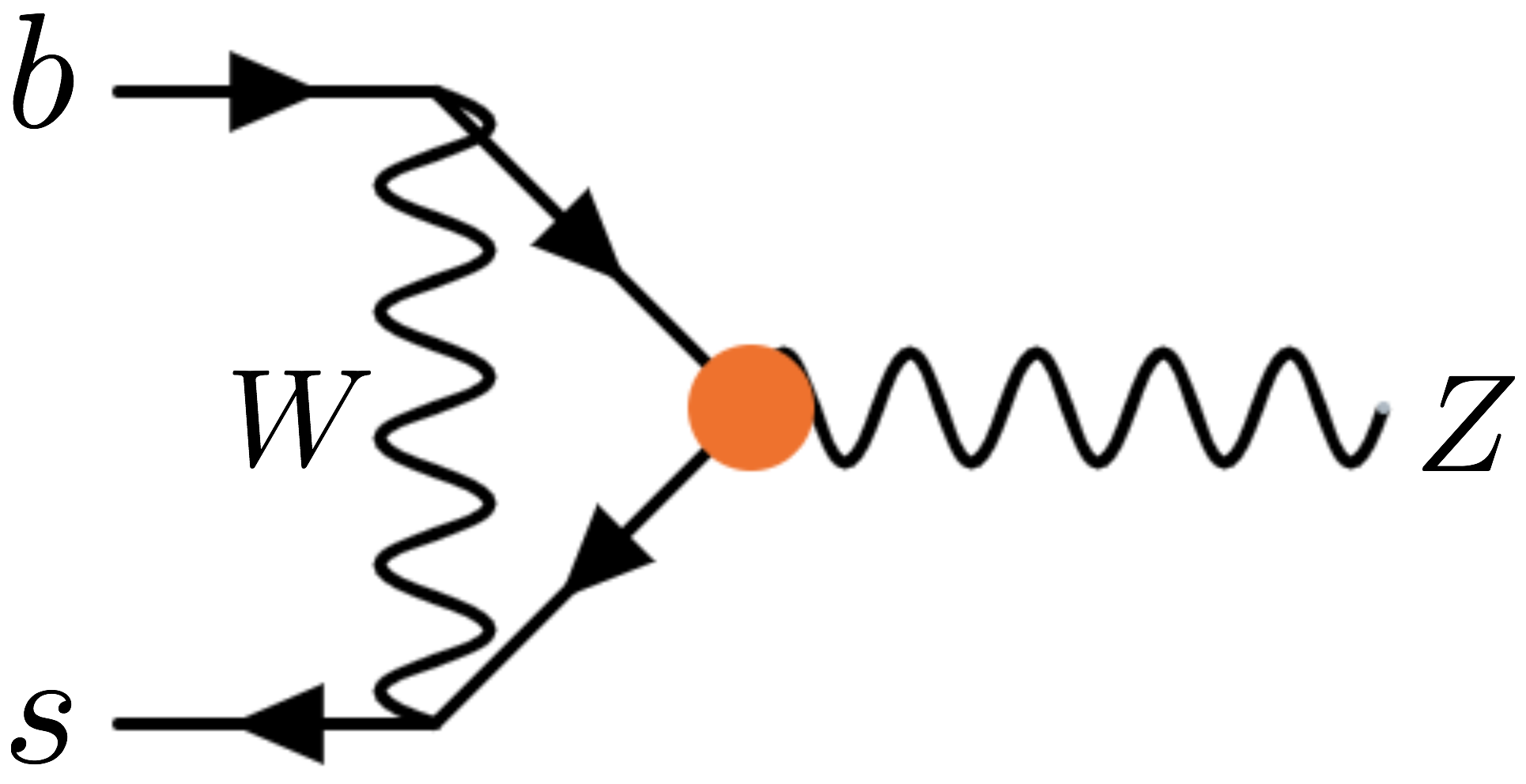}~~
\includegraphics[height=1.5cm]{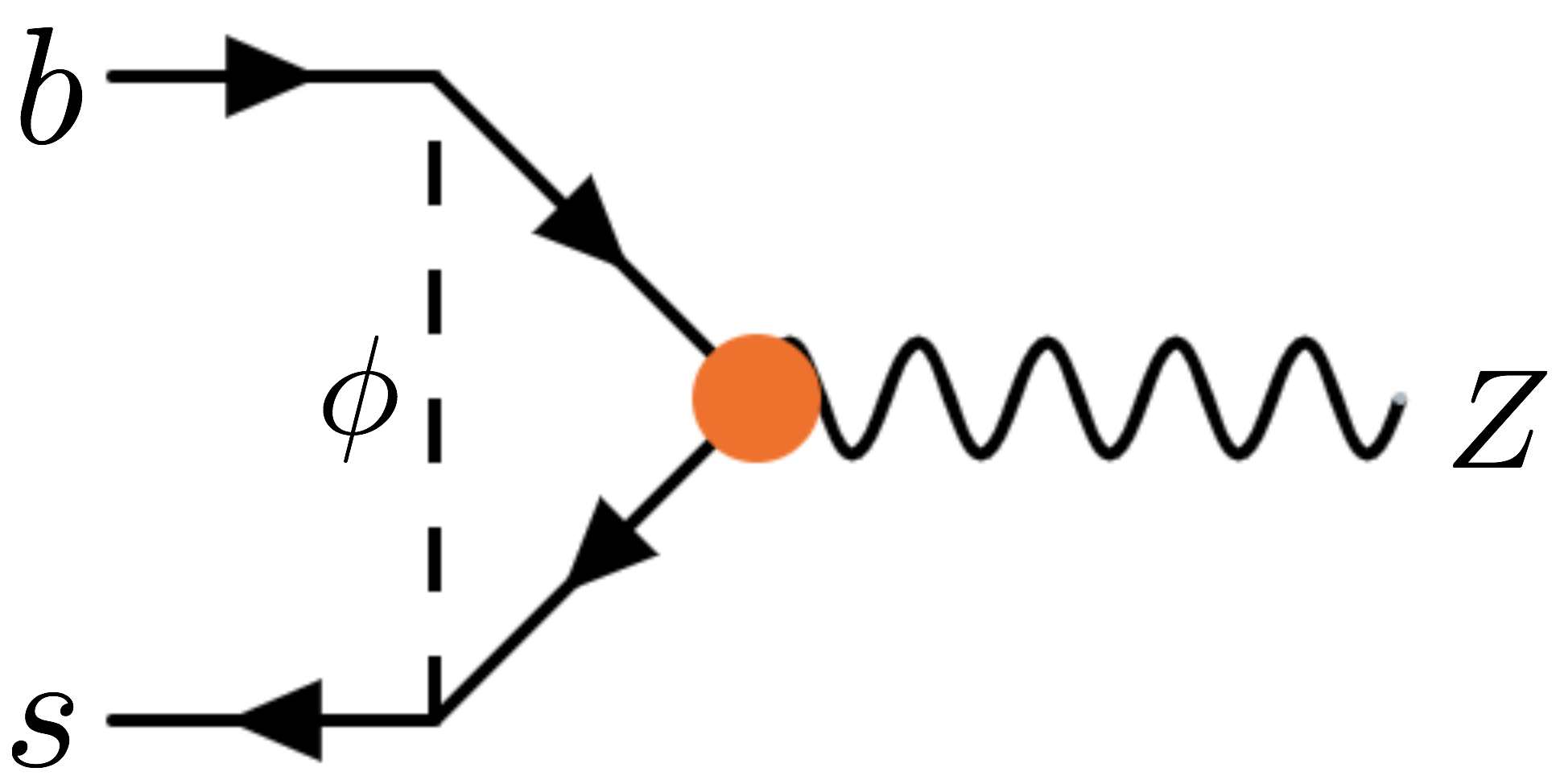}~~
\includegraphics[height=1.6cm]{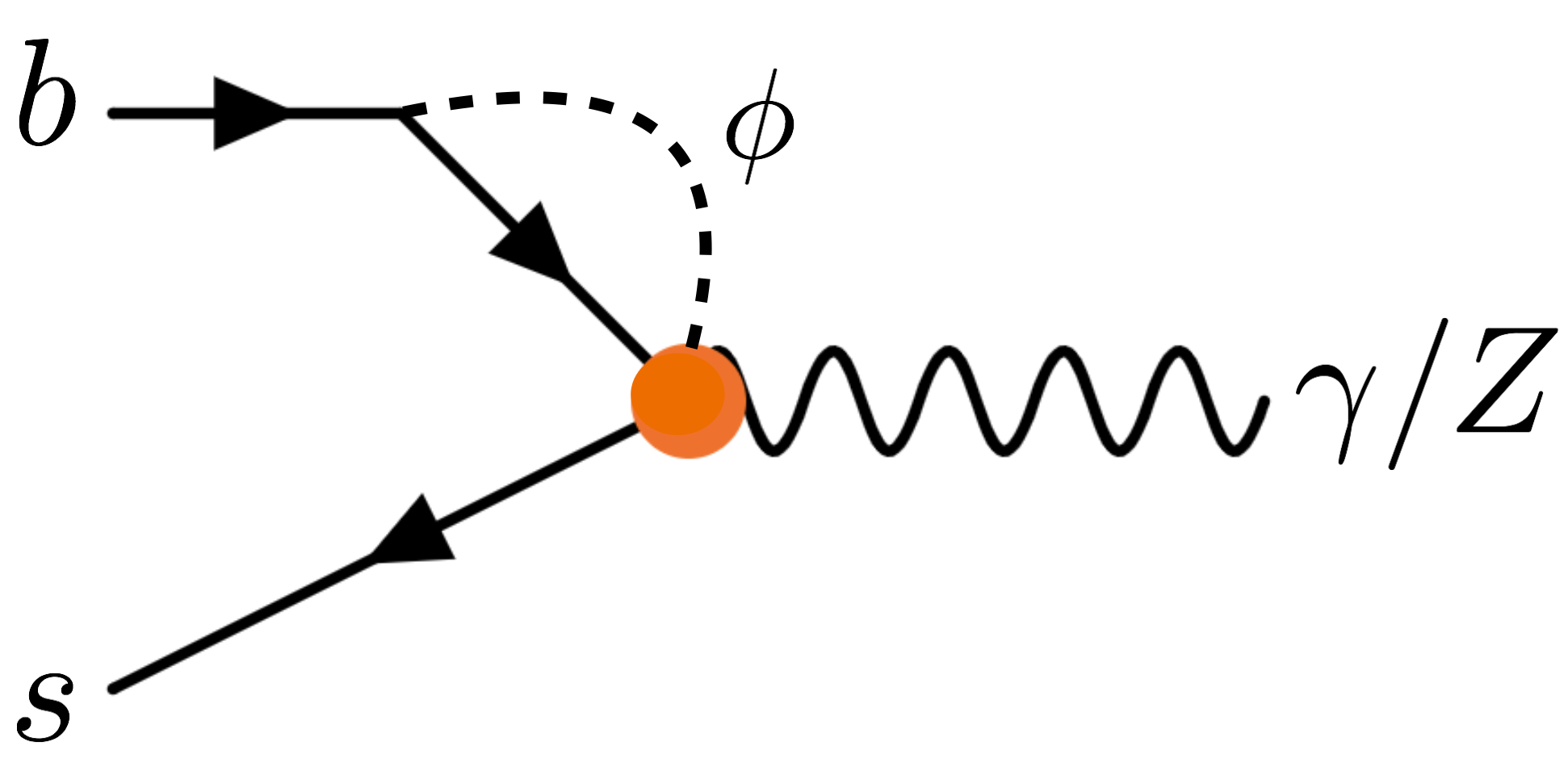}\\
\vspace{0.7cm}
\includegraphics[height=2cm]{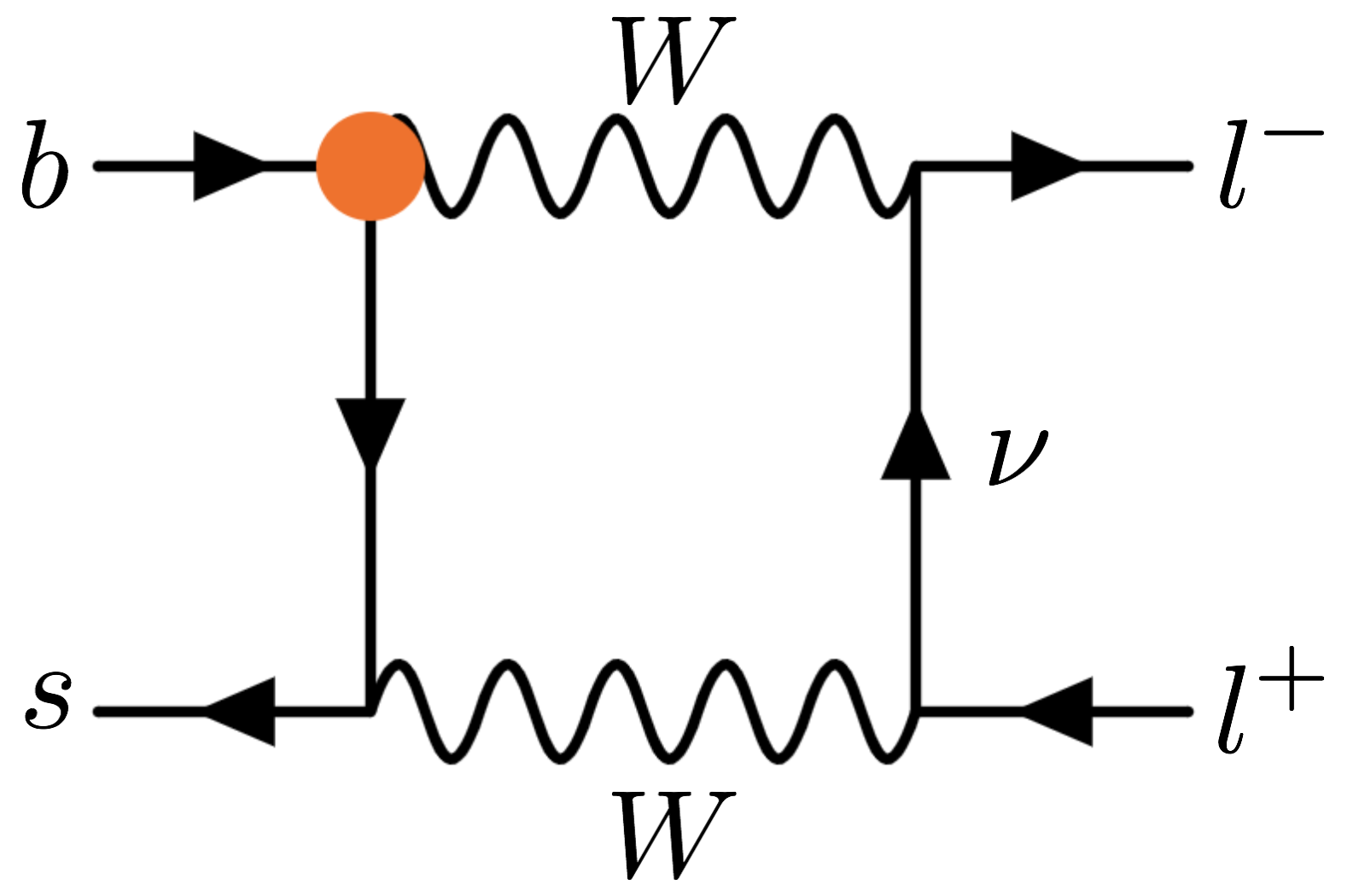}
\caption{Diagrams generating contributions to $b\to s l^+ l^-$ and/or $b\to s \gamma$ from the $Q_{Hq}^{(3)}$ operator. Diagrams in which the operator attaches to the $b$ quark leg imply also the existence (and inclusion in our calculations) of similar diagrams with the operator attached to the $s$ quark leg.  \label{CHq3}}
\end{center}
\end{figure}
This operator generates contributions to $B_s$ mixing via the diagrams in Fig.~\ref{CHq3mix}, and to $b\to s \gamma$ and $b\to s l^+ l^-$ via the diagrams in Fig.~\ref{CHq3}. There are also contributions to the chromomagnetic dipole operator via graphs similar to the second and fourth diagrams in Fig.~\ref{CHq3}, with a gluon replacing the photon. The Wilson coefficients of the mixing operator and the (chromo)magnetic dipole operators are simple scalings of the SM result:
\begin{align}
C_{1,\text{mix}}^s(x_t)&=4 v^2\,C_{Hq}^{(3)}\, S_0(x_t),\\
C_{1,\text{mix}}^s(x_c)&=4 v^2\,C_{Hq}^{(3)}\, S_0(x_c),\\
C_{1,\text{mix}}^s(x_t,x_c)&=4 v^2\,C_{Hq}^{(3)}\, S_0(x_t,x_c),\\
C_7&= -v^2\,C_{Hq}^{(3)}\,D_0^\prime(x_t),\\
C_8 &=-v^2\,C_{Hq}^{(3)}\,E_0^\prime(x_t).
\end{align}
where
\begin{align}
D_0^\prime(x_t) &= \frac{8 x_t^3 + 5 x_t^2 - 7 x_t}{12 (x_t - 1)^3} + \frac{x_t^2 (2 - 3 x_t)}{2 (1 - x_t)^4} \log x_t,\label{eqn:ILD0prime}\\
E_0^\prime(x_t) &=\frac{x_t (x_t^2 - 5 x_t - 2)}{4 (x_t - 1)^3} + \frac{3}{2}\frac{x_t^2}{(x_t - 1)^4} \log x_t,\label{eqn:ILE0prime}\\
S_0(x_t)&=\frac{4x_t-11x_t^2+x_t^3}{4(1-x_t)^2}-\frac{3x_t^3}{2(1-x_t)^3}\log x_t, \label{eqn:ILS0prime}\\
S_0(x_c)&= x_c \label{eqn:ILS0xc}, \\
S_0(x_t,x_c)&=x_c \left( \log \frac{x_t}{x_c}- \frac{3x_t}{4(1-x_t)}-\frac{3x_t^2}{4(1-x_t)^2} \log x_t \right) \label{eqn:ILS0xtxc}
\end{align}
are the usual Inami Lim functions~\cite{Inami:1980fz}. These results are in agreement with Refs.~\cite{Drobnak:2011wj,Drobnak:2011aa}.\footnote{Our results are in fact twice theirs; however this is accounted for by a slight difference in operator flavour structure. They study an operator containing a $b$ quark (without corresponding contributions for the first two generations), and hence only half of the charged current vertices of these diagrams can be affected by the operator; by contrast in our flavour structure all charged current vertices can be affected.}
Due to the presence of additional non SM-like diagrams, $C_9$ and $C_{10}$ contain pieces that are not just scalings of the SM result:
\begin{align}
C_9&= 2v^2\,C_{Hq}^{(3)}\left(\frac{4s_{\theta}^2-1}{s_{\theta}^2}I^{Hq3}(x_t) -\frac{1}{s_{\theta}^2}B_0(x_t) -D_0(x_t)\right),\\
C_{10}&=2v^2C_{Hq}^{(3)}\, \frac{1}{s_{\theta}^2} \left(B_0(x_t)+I^{Hq3}(x_t)\right),
\end{align}
where
\begin{align}
I^{Hq3}(x_t)&= \frac{x_t}{32}\left[ -7\log \frac{m_W^2}{\mu^2}+\frac{x_t+33}{2(1-x_t)}-\frac{7x_t^2-2x_t+12}{(1-x_t)^2}\log x_t\right],
\end{align}
and 
\begin{align}
B_0(x_t)&= \frac{1}{4}\left[\frac{x_t}{1-x_t} +\frac{x_t}{(x_t-1)^2}\log x_t\right], \label{eqn:ILB0}\\
C_0(x_t)&= \frac{x_t}{8}\left[\frac{x_t-6}{x_t-1}+\frac{3x_t+2}{(x_t-1)^2}\log x_t \right],\label{eqn:ILC0}\\
D_0(x_t)&=-\frac{4}{9}\log x_t +\frac{-19x_t^3+25x_t^2}{36(x_t-1)^3}+\frac{x_t^2(5x_t^2-2x_t-6)}{18(x_t-1)^4}\log x_t  \label{eqn:ILD0}
\end{align}
are again the usual Inami Lim~\cite{Inami:1980fz} functions.\footnote{Note that the functions $I^{Hq3}(x_t)$, $B_0(x_t)$, $C_0(x_t)$ and $D_0(x_t)$ are individually gauge parameter $\xi$ dependent, and are given here in Feynman gauge. The overall result is of course $\xi$-independent.}

\subsection{\boldmath{$Q_W$}}

\begin{figure}
\begin{center}
        \includegraphics[height=1.7cm]{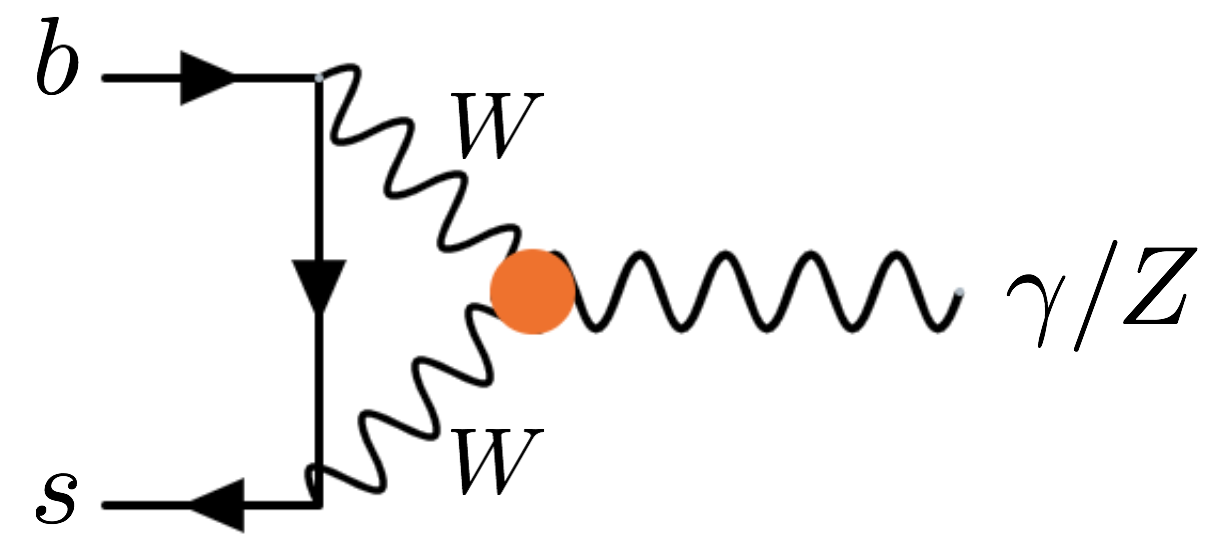}~~~
        \includegraphics[height=1.7cm]{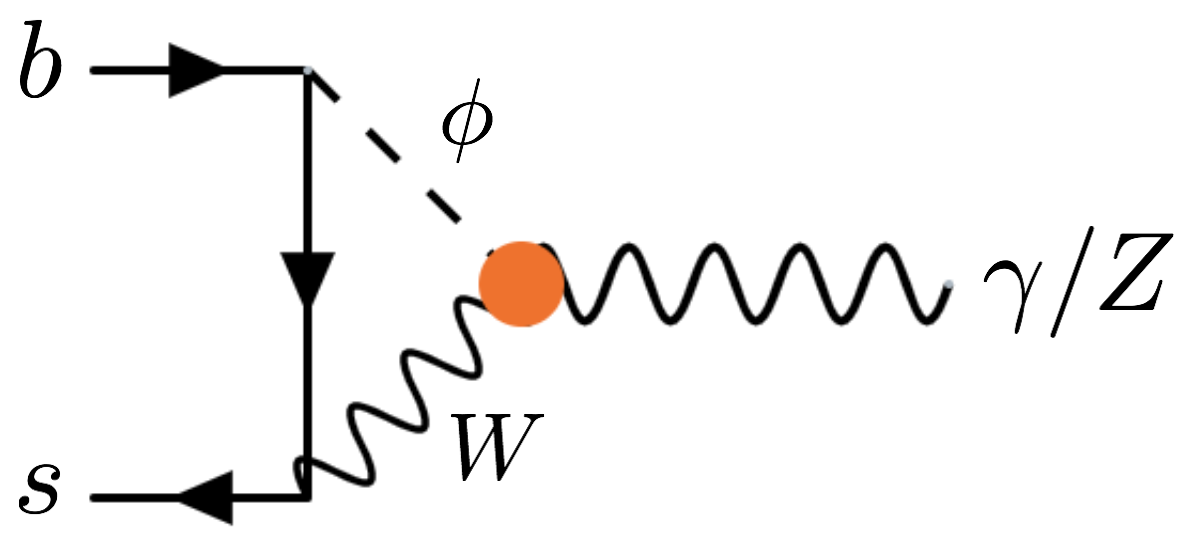}~~~
        \includegraphics[height=1.7cm]{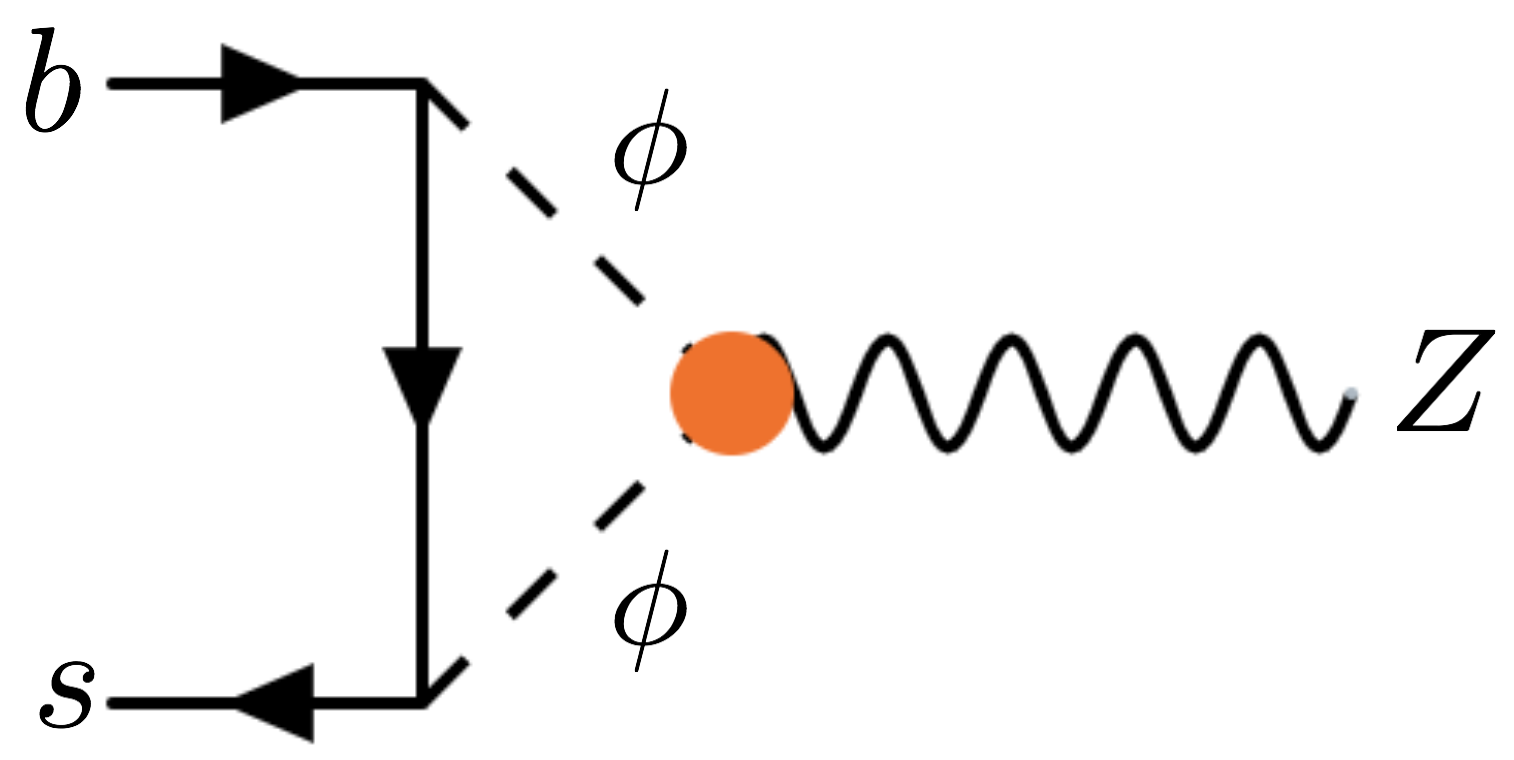}
\end{center}
\caption{Diagrams generating contributions to $b\to s l^+ l^-$ and/or $b\to s \gamma$ from $Q_W$ (first diagram only), $Q_{HD}$ (second and third diagrams only), and $Q_{HWB}$ (all three diagrams). Since the operators $Q_{HD}$ and $Q_{HWB}$ affect the $Z$ mass, and $Q_{HWB}$ affects the $Z$ and photon couplings by redefining the Weinberg angle, SM-like penguin diagrams also receive contributions from these operators and are included in our calculation. \label{CW}}
\end{figure}
This operator produces a triple gauge boson vertex with a different Lorentz structure to those in the SM, and generates contributions to $b\to s l^+ l^-$ and $b\to s \gamma$ via the first diagram in Fig.~\ref{CW}. These are
\begin{align}
C_7&=\frac{3}{2}g_2v^2C_W\left( -\frac{x_t^2+x_t}{2(x_t-1)^2} +\frac{x_t^2}{(x_t-1)^3}\log x_t\right),\\
C_9&=\frac{3}{2}g_2v^2C_W\, \left(\frac{3x_t^2-x_t}{2(x_t-1)^2} -\frac{x_t^3}{(x_t-1)^3}\log x_t \right).
\end{align}
Our result agrees with Ref.~\cite{Bobeth:2015zqa} (accounting for a difference in normalisation between their operator $O_{3W}$ and our operator $Q_W$).

\subsection{\boldmath{$Q_{HWB}$}}
This operator redefines the Weinberg angle (and hence enters $\gamma/Z$ vertices), but also induces new bosonic couplings with a different structure to those in the SM, directly generating contributions to $b\to s \gamma$ and $b\to s l^+ l^-$ via the diagrams in Fig.~\ref{CW}. In total we get
\begin{align}
C_7&=-v^2C_{HWB} \frac{g_2}{g_1}\left(\frac{8x_t^2-7x_t+5}{24(1-x_t)^3} +\frac{x_t(x_t^2-x_t+1)}{4(1-x_t)^4}\log x_t\right), \\
C_9 &=v^2C_{HWB}\frac{g_2}{g_1}\bigg(\frac{x_t(-9x_t^3+100x_t^2-178x_t+81}{18(1-x_t)^3} \nonumber \\&+\frac{39x_t^4-30x_t^3-81x_t^2+82x_t-16}{18(1-x_t)^4}\log x_t\bigg).
\end{align}

\subsection{\boldmath{$Q_{HD}$}}
The operator $Q_{HD}$ enters in the definition of theory parameters, notably $m_Z$ and the $Z$ couplings, due to its correction of the Higgs kinetic term; it also directly generates contributions to  $b\to s l^+ l^-$ via the last two diagrams in Fig.~\ref{CW}. In total, the contributions from this operator are
\begin{align}
C_7 &= \frac{1}{8}\frac{v^2}{s_{\theta}^2}C_{HD} (1-s_\theta^2)D_0^\prime(x_t), \\
C_9 &=\frac{1}{2} \frac{v^2}{s_{\theta}^2}C_{HD} \left[
(1-4s_\theta^2)I(x_t)+
(1-s_\theta^2)\left(D_0(x_t)+4 C_0(x_t) \right)\right],\\
C_{10}&=-\frac{1}{2}\frac{v^2}{s_{\theta}^2} C_{HD} I(x_t),
\end{align}
where $D_0^\prime(x_t)$, $C_0(x_t)$ and $D_0(x_t)$ are the usual Inami Lim functions defined in Eqns.~\eqref{eqn:ILD0prime}, \eqref{eqn:ILC0} and \eqref{eqn:ILD0}, and $I(x_t)$ has been defined in Eqn.~\eqref{eqn:I}.

\subsection{\boldmath{$Q_{Hl}^{(3)}$ and $Q_{ll}$}}
\begin{figure}
\begin{center}
\includegraphics[height=2cm]{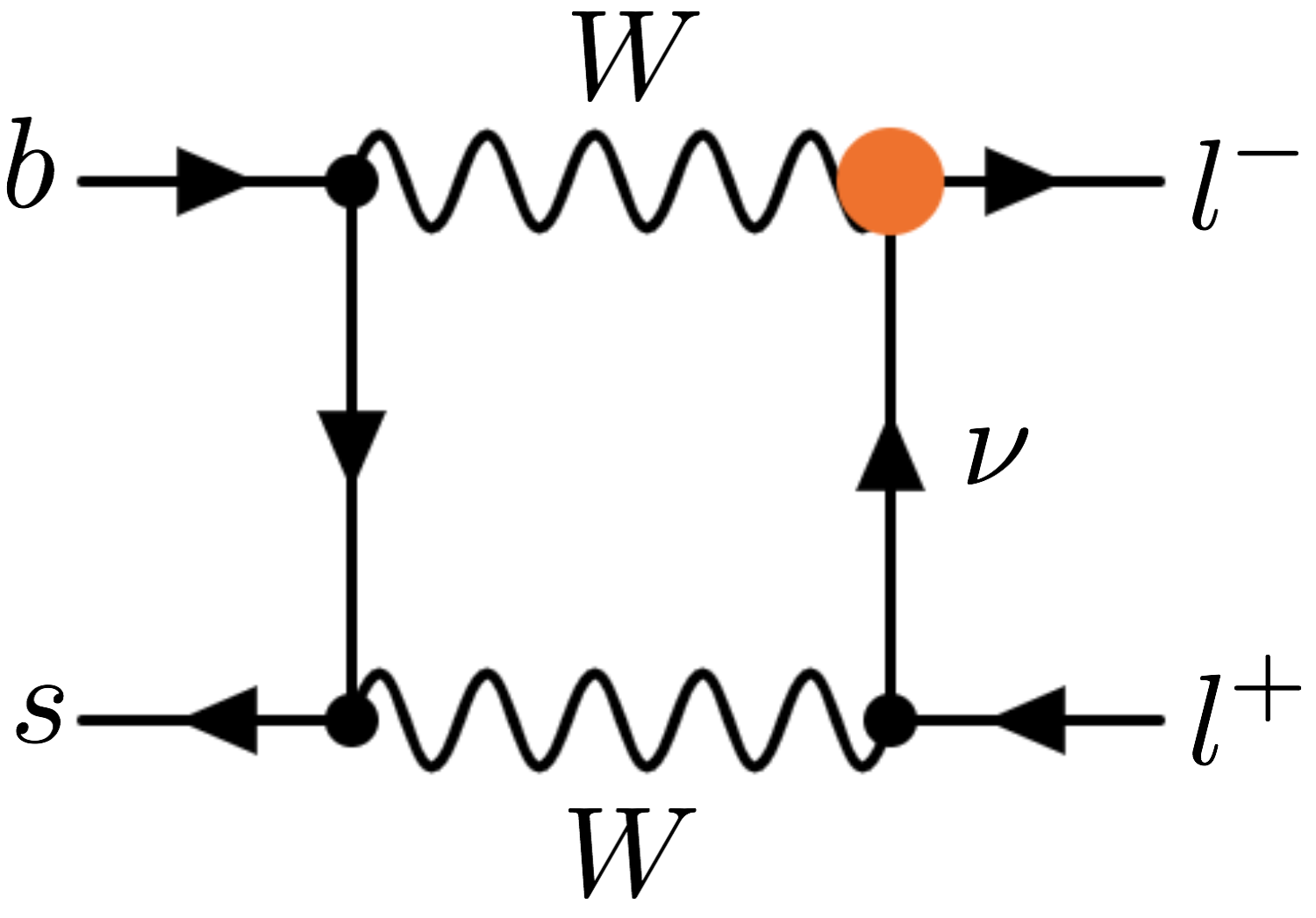}~~
\includegraphics[height=2cm]{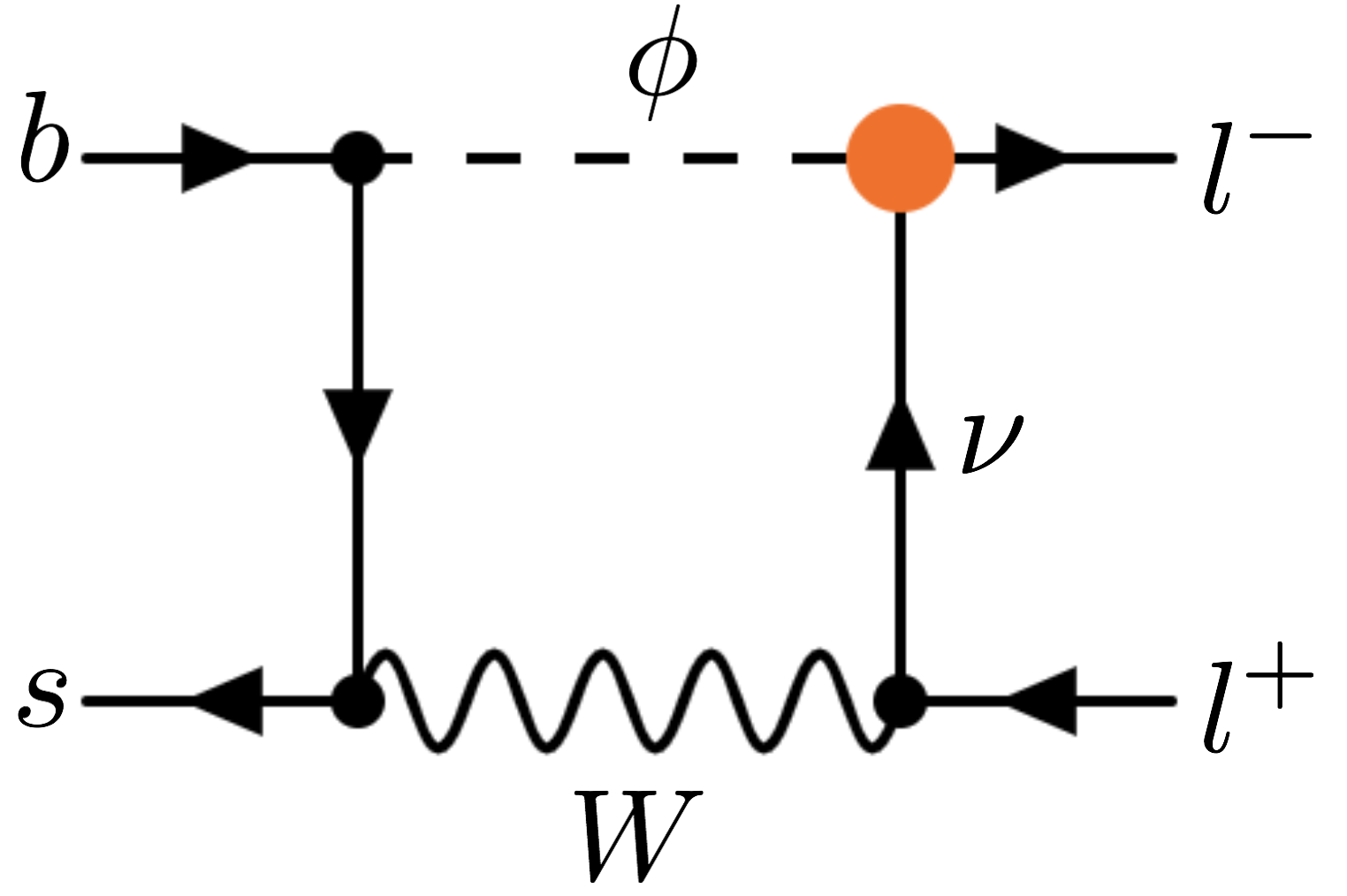}~~
\includegraphics[height=2cm]{newfigs/CHe5.png}~~
\includegraphics[height=1.8cm]{newfigs/leptonZpenguin.png}
\caption{Diagrams generating contributions to $b\to s l^+ l^-$ from the operator $Q_{Hl}^{(3)}$.  The first two diagrams should be taken to include also similar diagrams with the effective operator attaching to the $l^+$ line rather than the $l^-$ line, while the fourth diagram should be taken to include all other $Z$ penguin diagrams (including those with self-energies on external legs) where these operators affect the $Zl^+ l^-$ vertex.
\label{CHl3}}
\end{center}
\end{figure}

These operators enter every electroweak process since they are involved in the definition of $G_F$, which we take as an input parameter.  The $Q_{Hl}^{(3)}$ operator also generates direct contributions to the $b \to s l^+ l^-$ process, as shown in Fig.~\ref{CHl3}. 
As described in Sec.~\ref{sec:flavoursymm}, there are two ways of contracting the lepton doublets within $Q_{ll}$ to make a flavour singlet, so we can define two independent Wilson coefficients, $C_{ll}$ and $C_{ll}^{\prime}$, within our $U(3)^5$ invariant theory, although only $C_{ll}^{\prime}$ contributes here. 
Then the matching results for these operators are
\begin{align}
C_{1,\text{mix}}^s(x_t) &=-4v^2 \left( C_{Hl}^{(3)}-\frac{1}{2}C_{ll}^\prime \right) S_0(x_t),\\
C_{1,\text{mix}}^s(x_c) &=-4v^2 \left( C_{Hl}^{(3)}-\frac{1}{2}C_{ll}^\prime \right) S_0(x_c),\\
C_{1,\text{mix}}^s(x_t,x_c) &=-4v^2 \left( C_{Hl}^{(3)}-\frac{1}{2}C_{ll}^\prime \right) S_0(x_t,x_c),
\end{align}
\begin{align}
C_7 &= \frac{3}{2}v^2\left( C_{Hl}^{(3)}-\frac{1}{2}C_{ll}^\prime \right) D_0^\prime (x_t), \\
C_8 &= v^2 \left( C_{Hl}^{(3)}-\frac{1}{2}C_{ll}^\prime \right) E_0^\prime (x_t),\\
C_9 &=-4v^2 \left(C_{Hl}^{(3)}-\frac{1}{2}C_{ll}^\prime\right) \left( \frac{(1-4s_\theta^2)}{s_\theta^2}C_0(x_t)-\frac{1}{s_\theta^2}B_0(x_t)-D_0(x_t)\right)\nonumber \\
&- \frac{v^2}{s_{\theta}^2}C_{Hl}^{(3)}I^{Hl3}(x_t),\\
C_{10}&=-4v^2\, \frac{1}{s_\theta^2} \left(C_{Hl}^{(3)}-\frac{1}{2}C_{ll}^\prime \right)\left( B_0(x_t) - C_0(x_t) \right)+\frac{v^2}{s_{\theta}^2}C_{Hl}^{(3)}I^{Hl3}(x_t),
\end{align}
where
\begin{align}
I^{Hl3}(x_t)&= \frac{x_t}{16}\left[ \log \frac{m_W^2}{\mu^2}+\frac{7x_t-25}{2(1-x_t)}+\frac{x_t^2-14x_t+4}{(1-x_t)^2}\log x_t\right],
\end{align}
and the Inami Lim functions $B_0(x_t)$, $C_0(x_t)$, $D_0(x_t)$, $S_0(x_t)$, $S_0(x_c)$, $S_0(x_t,x_c)$, $D_0^\prime (x_t)$ and $E_0^\prime (x_t)$ were defined in Eqns.~\eqref{eqn:ILD0prime} -- \eqref{eqn:ILS0xtxc} and \eqref{eqn:ILB0} -- \eqref{eqn:ILD0}.

\section{Conclusions and outlook}
\label{sec:conc}

We have calculated here for the first time the full tree and one-loop matching of the SMEFT in the limit of complete $U(3)^5$ flavour symmetry in the UV onto the operators of the WET which contribute to down-sector FCNCs. Having these available, it will now be straightforward to run them to the appropriate experimental scale within the WET\footnote{Anomalous dimensions can be found in Refs.~\cite{Jenkins:2017dyc,Aebischer:2017gaw}} and derive the constraints that flavour observables provide on this flavour-symmetric limit of the SMEFT. In a forthcoming publication, we shall explore these constraints and compare their sensitivity to those arising from other measurements sensitive to similar sets of Wilson coefficients, for example electroweak precision observables, including the effects of operators which are allowed at linear order in the MFV expansion.  It is likely that these calculations will allow the derivation of constraints which lie in directions in parameter space linearly independent from those that currently exist from global fits to the flavour symmetric SMEFT. 

Ultimately, only a truly global picture of the SMEFT parameter space can give us meaningful insight into the underlying structure of new physics. If hints of physics beyond the SM appear in upcoming experiments, an understanding of global constraints on effective operators will signpost the most promising explanations, while if no deviation from SM predictions is seen this global picture will allow constraints to be easily placed on as-yet unimagined new models. The absence of explicit flavour structure is largely required if new physics is to be accessible at near-future experiments, or is expected to play a meaningful role in resolving the gauge hierarchy problem; this work provides tools to put new constraints from flavour observables on these models.

\acknowledgments

We gratefully acknowledge multiple helpful exchanges with Ilaria Brivio and Christoph Greub. The research of WS was partially funded by the Alexander von Humboldt Foundation in the framework of the Sofja Kovalevskaja Award 2016, endowed by the German Federal Ministry of Education and Research. 
This work also was supported by the Cluster of Excellence `Precision Physics, Fundamental Interactions, and Structure of Matter' (PRISMA+ EXC 2118/1) funded by the German Research Foundation (DFG) within the German Excellence Strategy (Project ID 39083149), as well as BMBF Verbundprojekt 05H2018 - Belle II: Indirekte Suche nach neuer Physik bei Belle-II. TH thanks the CERN theory group for its hospitality during his regular visits to CERN where part of the work was done. SR is grateful to UCSB physics department and KITP Santa Barbara for hospitality while working on this.

\bibliography{FCNCbib}

\appendix

\section{Input parameter dependence of results}
\label{sec:inputsappendix}

As described in Sec.~\ref{sec:methods}, there is a choice to be made in the electroweak input parameters, which affects the pieces of the result which depend on the resulting dimension-six input parameter shifts $\delta g_1$, $\delta g_2$ and $\delta v$. In the main text we have chosen to take the three measured electroweak input parameters as $\lbrace \hat{m}_W, \hat{m}_Z, \hat{G}_F \rbrace$. The purpose of this Appendix is to clarify which pieces of our results are dependent on this choice, and to provide results for the alternative input scheme in which $\lbrace \hat{\alpha}_{em}, \hat{m}_Z, \hat{G}_F\rbrace$ are the set of measured electroweak inputs.

In the $\lbrace\hat{m}_W, \hat{m}_Z, \hat{G}_F \rbrace$ input scheme, the basis shifts are given by~\cite{Brivio:2017btx}
\begin{align}
\frac{\delta g_1}{g_1} &=v^2\left(\left(-C_{Hl}^{(3)}+\frac{1}{2}C_{ll}^\prime\right)-\frac{1}{4g_1^2}(g_2^2+g_1^2)C_{HD}-\frac{g_2}{g_1} C_{HWB}\right),\\
\frac{\delta g_2}{g_2} &=-v^2 \left(-\frac{1}{2}C_{ll}^\prime+C_{Hl}^{(3)}\right ), \\
\frac{\delta v}{v}&=v^2\left(-\frac{1}{2}C_{ll}^\prime+C_{Hl}^{(3)}\right),
\end{align}
while in the $\left\lbrace \hat{\alpha}_{em}, \hat{m}_Z, \hat{G}_F \right \rbrace$ input scheme, the basis shifts are defined
\begin{align}
\frac{\delta g_1}{g_1} &=-\frac{g_1 v^2}{(g_1^2-g_2^2)}\left(\frac{g_1}{4}C_{HD} +g_2 C_{HWB}\right),\\
\frac{\delta g_2}{g_2}  &= \frac{g_2 v^2}{(g_1^2-g_2^2)}\left(\frac{g_2}{4}C_{HD} +g_1 C_{HWB}\right),\\
\frac{\delta v}{v}&=v^2\left(-\frac{1}{2}C_{ll}^\prime+C_{Hl}^{(3)}\right).
\end{align}
For the definitions of hatted parameters in the $\lbrace \hat \alpha_{em}, \hat m_Z, \hat G_F \rbrace$ input scheme in terms of the measured inputs, we refer to Ref.~\cite{Brivio:2017btx}.
In order to allow switching between the different schemes, our results can be written as shift-independent pieces (i.e. pieces that remain when all three $\delta$s are set to zero) plus pieces written in terms of these three $\delta$s. We emphasise that our results in the main text in Sec.~\ref{sec:results} are the sum of these pieces, for the $\lbrace \hat{m}_W, \hat{m}_Z, \hat{G}_F \rbrace$ input scheme. The shift-independent pieces and the shift-dependent pieces of our calculation are separately independent of the gauge parameter $\xi$, as they must be since the gauge-invariance of the SMEFT does not rely on the values of the Lagrangian parameters $\bar{g}_1$, $\bar{g}_2$, and $\bar{v}$.

In the next subsection, we present the shift-independent pieces (meaning, specifically, the results you get by performing only steps 1 and 2 of the procedure in Sec.~\ref{sec:methods}) for the three operators that also appear in the input shifts. Then we present the extra shift-dependent pieces that should be added to these in the $\lbrace \hat m_W, \hat m_Z, \hat G_F \rbrace$ scheme (Sec.~\ref{sec:scheme1shifts}) or in the $\lbrace \hat{\alpha}_{em}, \hat{m}_Z, \hat{G}_F \rbrace$ scheme (Sec.~\ref{sec:scheme2shifts}). Note that the form of the corrections in the first scheme, chosen for our main results presentation, is notably simpler than that in the second; this is due to the fact that $m_W$ is a fundamental input in the first and a complicated derived quantity in the second, and $m_W$ appears very often throughout these calculations.

\subsection{Input shift independent pieces}
The only SMEFT Wilson coefficients which appear in both the input shift pieces and the shift-independent pieces, for the two schemes given above, are $C_{Hl}^{(3)}$, $C_{HD}$, and $C_{HWB}$. The coefficient $C_{ll}^\prime$ only contributes via input shifts. For all other Wilson coefficients, the results given in the main text are shift-independent.

\subsubsection{\boldmath{$Q_{Hl}^{(3)}$}}
\begin{align}
C_9&=-v^2 C_{Hl}^{(3)}\frac{1}{s_\theta^2}\frac{x_t}{16}\left(\log \frac{m_W^2}{\mu^2}+\frac{7x_t-25}{2(1-x_t)}+\frac{x_t^2-14x_t+4}{(1-x_t)^2}\log x_t \right),\\
C_{10} &=v^2 C_{Hl}^{(3)}\frac{1}{s_\theta^2}\frac{x_t}{16}\left(\log \frac{m_W^2}{\mu^2}+\frac{7x_t-25}{2(1-x_t)}+\frac{x_t^2-14x_t+4}{(1-x_t)^2}\log x_t \right).
\end{align}

\subsubsection{\boldmath{$Q_{HD}$}}
\begin{align}
C_9&=-\frac{1}{2}\frac{4s_\theta^2-1}{s_\theta^2}v^2 C_{HD} I(x_t),\\
C_{10} &=-\frac{1}{2}\frac{1}{s_\theta^2}v^2 C_{HD} I(x_t),
\end{align}
where $I(x_t)$ is defined in Eqn.~\eqref{eqn:I}.

\subsubsection{\boldmath{{$C_{HWB}$}}}
\begin{align}
C_7&=v^2C_{HWB}\frac{g_2}{g_1}\left(\frac{g_1^2-g_2^2}{g_1^2+g_2^2}\frac{1}{4}D_0^\prime (x_t)+I^{HWB}_7(x_t)\right),\\
C_9 &=v^2C_{HWB}\frac{g_2}{g_1}\left(\frac{g_1^2-g_2^2}{g_1^2+g_2^2}\left(4 C_0(x_t)+D_0(x_t) \right) +I^{HWB}_9(x_t)\right),
\end{align}
where
\begin{align}
I^{HWB}_{7}(x_t)&=-\frac{x_t(8x_t^2-19x_t+17)}{48(1-x_t)^3}-\frac{x_t(2x_t^2-5x_t+4)}{8(1-x_t)^4} \log x_t,\\
I^{HWB}_{9}(x_t)&=\frac{x_t(37x_t^2-97x_t+54)}{36(1-x_t)^3}+\frac{15x_t^4-24x_t^3-18x_t^2+32x_t-8}{18(1-x_t)^4} \log x_t,
\end{align}
and $C_0(x_t)$, $D_0(x_t)$ and $D_0^\prime (x_t)$ are the usual Inami Lim functions defined in Eqns.~\eqref{eqn:ILC0}, \eqref{eqn:ILD0} and \eqref{eqn:ILD0prime}.

\subsection{Input shift pieces for $\lbrace \hat{m}_W, \hat{m}_Z, \hat{G}_F \rbrace$ scheme}
\label{sec:scheme1shifts}

\begin{align}
C_{1,\text{mix}}^s(x_t) &=-4v^2 \left( C_{Hl}^{(3)}-\frac{1}{2}C_{ll}^\prime \right) S_0(x_t),\\
C_{1,\text{mix}}^s(x_c) &=-4v^2 \left( C_{Hl}^{(3)}-\frac{1}{2}C_{ll}^\prime \right) S_0(x_c),\\
C_{1,\text{mix}}^s(x_t,x_c) &=-4v^2 \left( C_{Hl}^{(3)}-\frac{1}{2}C_{ll}^\prime \right) S_0(x_t,x_c),\\
C_7 &=v^2 \left[\frac{3}{2}\left(C_{Hl}^{(3)}-\frac{1}{2}C_{ll}^\prime \right)+\frac{1}{2}\frac{g_2}{g_1}c_\theta^2 C_{HWB}+\frac{1}{8}\frac{g_2^2}{g_1^2}C_{HD} \right] D_0^\prime (x_t), \\
C_8 &=v^2  \left(C_{Hl}^{(3)}-\frac{1}{2}C_{ll}^\prime \right) E_0^\prime (x_t), \\
C_9 &=-4v^2 \left(C_{Hl}^{(3)}-\frac{1}{2}C_{ll}^\prime\right) \left( \frac{(1-4s_\theta^2)}{s_\theta^2}C_0(x_t)-\frac{1}{s_\theta^2}B_0(x_t)-D_0(x_t)\right) \nonumber \\
&+ v^2 \left( 2\,\frac{g_2}{g_1}c_\theta^2 C_{HWB}+\frac{1}{2}\frac{g_2^2}{g_1^2}C_{HD}\right) \left(D_0(x_t)+4 C_0(x_t) \right),\\
C_{10} &=-4v^2\, \frac{1}{s_\theta^2} \left(C_{Hl}^{(3)}-\frac{1}{2}C_{ll}^\prime\right)\left( B_0(x_t) - C_0(x_t) \right),
\end{align}
where $B_0(x_t)$, $C_0(x_t)$, $D_0(x_t)$, $D_0^\prime (x_t)$, $E_0^\prime (x_t)$ $S_0 (x_t)$, $S_0 (x_c)$ and $S_0 (x_t,x_c)$ are the usual Inami Lim functions defined in Eqns.~\eqref{eqn:ILD0prime} -- \eqref{eqn:ILS0xtxc} and \eqref{eqn:ILB0} -- \eqref{eqn:ILD0}.

\subsection{Input shift pieces for $\lbrace \hat{\alpha}_{em}, \hat{m}_Z, \hat{G}_F\rbrace$ scheme}
\label{sec:scheme2shifts}
\begin{align}
C_{1,\text{mix}}^s(x_t) &=-4v^2 \left( C_{Hl}^{(3)}-\frac{1}{2}C_{ll}^\prime \right) S_0(x_t),\\
C_{1,\text{mix}}^s(x_c) &=-4v^2 \left( C_{Hl}^{(3)}-\frac{1}{2}C_{ll}^\prime \right) S_0(x_c),\\
C_{1,\text{mix}}^s(x_t,x_c) &=-4v^2 \left( C_{Hl}^{(3)}-\frac{1}{2}C_{ll}^\prime \right) S_0(x_t,x_c),
\end{align}
\begin{align}
C_7 &= v^2\left( C_{Hl}^{(3)}-\frac{1}{2}C_{ll}^\prime\right)\left[ \frac{x_t(2x_t^3-75x_t^2-66x_t+67)}{72(1-x_t)^4}-\frac{x_t(3x_t^3+19x_t^2-6x_t-4)}{12(1-x_t)^5}\log x_t \right]\nonumber \\
&-\frac{g_2^2v^2}{(g_1^2-g_2^2)}\left(\frac{1}{4}C_{HD}+\frac{g_1}{g_2}C_{HWB}\right)\bigg[\frac{x_t(46x_t^3+57x_t^2-6x_t-25)}{72(1-x_t)^4}\nonumber\\
&+\frac{x_t(21x_t^3-11x_t^2+6x_t-4)}{12(1-x_t)^5}\log x_t \bigg]-\frac{1}{2}\frac{g_1 g_2 v^2}{(g_1^2+g_2^2)}C_{HWB} D_0^\prime (x_t),
\end{align}
\begin{align}
C_8 &=v^2\left( C_{Hl}^{(3)}-\frac{1}{2}C_{ll}^\prime\right)\left[ -\frac{x_t(x_t^3-6x_t^2-15x_t-16)}{12(1-x_t)^4}+\frac{x_t(2x_t^2+3x_t+1)}{2(1-x_t)^5}\log x_t \right]\nonumber\\
&-\frac{g_2^2v^2}{(g_1^2-g_2^2)}\left(\frac{1}{4}C_{HD}+\frac{g_1}{g_2}C_{HWB}\right)\left[\frac{x_t(2x_t^3-12x_t^2-3x_t-5)}{6(1-x_t)^4} -\frac{x_t(5x_t^2+1)}{2(1-x_t)^5}\log x_t \right],
\end{align}
\begin{align}
C_9 &= v^2\left( C_{Hl}^{(3)}-\frac{1}{2}C_{ll}^\prime\right)\bigg[-\frac{1}{s_\theta^2}\left( \frac{x_t(x_t^2-2x_t+4)}{2(1-x_t)^2}-\frac{3x_t^2(x_t-3)}{4(1-x_t)^3}\log x_t \right)\nonumber\\
&+\frac{x_t(108x_t^4-530x_t^3+1557x_t^2-1908x_t+737)}{54(1-x_t)^4}\nonumber\\&-\frac{24x_t^5-126x_t^4+123x_t^3+35x_t^2-54x_t+4}{9(1-x_t)^5}\log x\bigg] \nonumber \\
&+\frac{g_2^2v^2}{(g_1^2-g_2^2)}\left(\frac{1}{4}C_{HD}+\frac{g_1}{g_2}C_{HWB}\right)\bigg[\frac{x_t(54x_t^4-149x_t^3+615x_t^2-969x_t+413)}{54(1-x_t)^4}\nonumber\\
& -\frac{27x_t^5-219x_t^4+333x_t^3-129x_t^2+8x_t-8}{18(1-x_t)^5} \log x_t \bigg]\nonumber \\
&+\frac{g_2^2}{g_1^2}v^2 \left(\frac{1}{4}C_{HD}+\frac{g_1}{g_2}C_{HWB}\right) \left(\frac{3x_t^2}{2(1-x_t)^2}+\frac{3x_t^2(1+x_t)}{4(1-x_t)^3}\log x_t \right)\nonumber \\
&- 2\frac{g_1 g_2 v^2}{g_1^2+g_2^2}C_{HWB} \left(4C_0(x_t)+D_0(x_t) \right) ,
\end{align}
\begin{align}
C_{10} &= \frac{v^2}{s_\theta^2}\left( C_{Hl}^{(3)}-\frac{1}{2}C_{ll}^\prime\right)\left[ \frac{x_t(x_t^2-2x_t+4)}{2(1-x_t)^2}-\frac{3x_t^2(x_t-3)}{4(1-x_t)^3}\log x_t \right]\nonumber\\
&+\frac{1}{s_\theta^2}\frac{g_2^2v^2}{(g_1^2-g_2^2)}\left(\frac{1}{4}C_{HD}+\frac{g_1}{g_2}C_{HWB}\right)\left[\frac{3x_t^2}{2(1-x_t)^2}+\frac{3x_t^2(x_t+1)}{4(1-x_t)^3}\log x_t \right],
\end{align}
where $B_0(x_t)$, $C_0(x_t)$, $D_0(x_t)$, $D_0^\prime (x_t)$, $E_0^\prime (x_t)$, $S_0 (x_t)$, $S_0 (x_c)$ and $S_0 (x_t,x_c)$ are the usual Inami Lim functions defined in Eqns.~\eqref{eqn:ILD0prime} -- \eqref{eqn:ILS0xtxc} and \eqref{eqn:ILB0} -- \eqref{eqn:ILD0}.

\begin{table}
\begin{center}
\small
\begin{minipage}[t]{4.4cm}
\renewcommand{\arraystretch}{1.5}
\begin{tabular}[t]{c|c}
\multicolumn{2}{c}{$1:X^3$} \\
\hline
$Q_G$                & $f^{ABC} G_\mu^{A\nu} G_\nu^{B\rho} G_\rho^{C\mu} $ \\
$Q_{\widetilde G}$          & $f^{ABC} \widetilde G_\mu^{A\nu} G_\nu^{B\rho} G_\rho^{C\mu} $ \\
$Q_W$                & $\epsilon^{IJK} W_\mu^{I\nu} W_\nu^{J\rho} W_\rho^{K\mu}$ \\ 
$Q_{\widetilde W}$          & $\epsilon^{IJK} \widetilde W_\mu^{I\nu} W_\nu^{J\rho} W_\rho^{K\mu}$ \\
\end{tabular}
\end{minipage}
\begin{minipage}[t]{2.6cm}
\renewcommand{\arraystretch}{1.5}
\begin{tabular}[t]{c|c}
\multicolumn{2}{c}{$2:H^6$} \\
\hline
$Q_H$       & $(H^\dag H)^3$ 
\end{tabular}
\end{minipage}
\begin{minipage}[t]{5.1cm}
\renewcommand{\arraystretch}{1.5}
\begin{tabular}[t]{c|c}
\multicolumn{2}{c}{$3:H^4 D^2$} \\
\hline
$Q_{H\Box}$ & $(H^\dag H)\Box(H^\dag H)$ \\
$Q_{H D}$   & $\ \left(H^\dag D_\mu H\right)^* \left(H^\dag D_\mu H\right)$ 
\end{tabular}
\end{minipage}
\begin{minipage}[t]{2.5cm}

\renewcommand{\arraystretch}{1.5}
\begin{tabular}[t]{c|c}
\multicolumn{2}{c}{$5: \psi^2H^3 + \hbox{h.c.}$} \\
\hline
$Q_{eH}$           & $(H^\dag H)(\bar l_p e_r H)$ \\
$Q_{uH}$          & $(H^\dag H)(\bar q_p u_r \widetilde H )$ \\
$Q_{dH}$           & $(H^\dag H)(\bar q_p d_r H)$\\
\end{tabular}
\end{minipage}

\vspace{0.25cm}

\begin{minipage}[t]{4.6cm}
\renewcommand{\arraystretch}{1.5}
\begin{tabular}[t]{c|c}
\multicolumn{2}{c}{$4:X^2H^2$} \\
\hline
$Q_{H G}$     & $H^\dag H\, G^A_{\mu\nu} G^{A\mu\nu}$ \\
$Q_{H\widetilde G}$         & $H^\dag H\, \widetilde G^A_{\mu\nu} G^{A\mu\nu}$ \\
$Q_{H W}$     & $H^\dag H\, W^I_{\mu\nu} W^{I\mu\nu}$ \\
$Q_{H\widetilde W}$         & $H^\dag H\, \widetilde W^I_{\mu\nu} W^{I\mu\nu}$ \\
$Q_{H B}$     & $ H^\dag H\, B_{\mu\nu} B^{\mu\nu}$ \\
$Q_{H\widetilde B}$         & $H^\dag H\, \widetilde B_{\mu\nu} B^{\mu\nu}$ \\
$Q_{H WB}$     & $ H^\dag \tau^I H\, W^I_{\mu\nu} B^{\mu\nu}$ \\
$Q_{H\widetilde W B}$         & $H^\dag \tau^I H\, \widetilde W^I_{\mu\nu} B^{\mu\nu}$ 
\end{tabular}
\end{minipage}
\begin{minipage}[t]{5.1cm}
\renewcommand{\arraystretch}{1.5}
\begin{tabular}[t]{c|c}
\multicolumn{2}{c}{$6:\psi^2 XH+\hbox{h.c.}$} \\
\hline
$Q_{eW}$      & $(\bar l_p \sigma^{\mu\nu} e_r) \tau^I H W_{\mu\nu}^I$ \\
$Q_{eB}$        & $(\bar l_p \sigma^{\mu\nu} e_r) H B_{\mu\nu}$ \\
$Q_{uG}$        & $(\bar q_p \sigma^{\mu\nu} T^A u_r) \widetilde H \, G_{\mu\nu}^A$ \\
$Q_{uW}$        & $(\bar q_p \sigma^{\mu\nu} u_r) \tau^I \widetilde H \, W_{\mu\nu}^I$ \\
$Q_{uB}$        & $(\bar q_p \sigma^{\mu\nu} u_r) \widetilde H \, B_{\mu\nu}$ \\
$Q_{dG}$        & $(\bar q_p \sigma^{\mu\nu} T^A d_r) H\, G_{\mu\nu}^A$ \\
$Q_{dW}$         & $(\bar q_p \sigma^{\mu\nu} d_r) \tau^I H\, W_{\mu\nu}^I$ \\
$Q_{dB}$        & $(\bar q_p \sigma^{\mu\nu} d_r) H\, B_{\mu\nu}$ 
\end{tabular}
\end{minipage}
\begin{minipage}[t]{5.2cm}
\renewcommand{\arraystretch}{1.5}
\begin{tabular}[t]{c|c}
\multicolumn{2}{c}{$7:\psi^2H^2 D$} \\
\hline
$Q_{H l}^{(1)}$      & $(H^\dag i\overleftrightarrow{D}_\mu H)(\bar l_p \gamma^\mu l_r)$\\
$Q_{H l}^{(3)}$      & $(H^\dag i\overleftrightarrow{D}^I_\mu H)(\bar l_p \tau^I \gamma^\mu l_r)$\\
$Q_{H e}$            & $(H^\dag i\overleftrightarrow{D}_\mu H)(\bar e_p \gamma^\mu e_r)$\\
$Q_{H q}^{(1)}$      & $(H^\dag i\overleftrightarrow{D}_\mu H)(\bar q_p \gamma^\mu q_r)$\\
$Q_{H q}^{(3)}$      & $(H^\dag i\overleftrightarrow{D}^I_\mu H)(\bar q_p \tau^I \gamma^\mu q_r)$\\
$Q_{H u}$            & $(H^\dag i\overleftrightarrow{D}_\mu H)(\bar u_p \gamma^\mu u_r)$\\
$Q_{H d}$            & $(H^\dag i\overleftrightarrow{D}_\mu H)(\bar d_p \gamma^\mu d_r)$\\
$Q_{H u d}$ + h.c.   & $i(\widetilde H ^\dag D_\mu H)(\bar u_p \gamma^\mu d_r)$\\
\end{tabular}
\end{minipage}

\vspace{0.25cm}

\begin{minipage}[t]{4.75cm}
\renewcommand{\arraystretch}{1.5}
\begin{tabular}[t]{c|c}
\multicolumn{2}{c}{$8:(\bar LL)(\bar LL)$} \\
\hline
$Q_{\ell \ell}$        & $(\bar l_p \gamma_\mu l_r)(\bar l_s \gamma^\mu l_t)$ \\
$Q_{qq}^{(1)}$  & $(\bar q_p \gamma_\mu q_r)(\bar q_s \gamma^\mu q_t)$ \\
$Q_{qq}^{(3)}$  & $(\bar q_p \gamma_\mu \tau^I q_r)(\bar q_s \gamma^\mu \tau^I q_t)$ \\
$Q_{\ell q}^{(1)}$                & $(\bar l_p \gamma_\mu l_r)(\bar q_s \gamma^\mu q_t)$ \\
$Q_{\ell q}^{(3)}$                & $(\bar l_p \gamma_\mu \tau^I l_r)(\bar q_s \gamma^\mu \tau^I q_t)$ 
\end{tabular}
\end{minipage}
\begin{minipage}[t]{5.25cm}
\renewcommand{\arraystretch}{1.5}
\begin{tabular}[t]{c|c}
\multicolumn{2}{c}{$8:(\bar RR)(\bar RR)$} \\
\hline
$Q_{ee}$               & $(\bar e_p \gamma_\mu e_r)(\bar e_s \gamma^\mu e_t)$ \\
$Q_{uu}$        & $(\bar u_p \gamma_\mu u_r)(\bar u_s \gamma^\mu u_t)$ \\
$Q_{dd}$        & $(\bar d_p \gamma_\mu d_r)(\bar d_s \gamma^\mu d_t)$ \\
$Q_{eu}$                      & $(\bar e_p \gamma_\mu e_r)(\bar u_s \gamma^\mu u_t)$ \\
$Q_{ed}$                      & $(\bar e_p \gamma_\mu e_r)(\bar d_s\gamma^\mu d_t)$ \\
$Q_{ud}^{(1)}$                & $(\bar u_p \gamma_\mu u_r)(\bar d_s \gamma^\mu d_t)$ \\
$Q_{ud}^{(8)}$                & $(\bar u_p \gamma_\mu T^A u_r)(\bar d_s \gamma^\mu T^A d_t)$ \\
\end{tabular}
\end{minipage}
\begin{minipage}[t]{4.75cm}
\renewcommand{\arraystretch}{1.5}
\begin{tabular}[t]{c|c}
\multicolumn{2}{c}{$8:(\bar LL)(\bar RR)$} \\
\hline
$Q_{le}$               & $(\bar l_p \gamma_\mu l_r)(\bar e_s \gamma^\mu e_t)$ \\
$Q_{lu}$               & $(\bar l_p \gamma_\mu l_r)(\bar u_s \gamma^\mu u_t)$ \\
$Q_{ld}$               & $(\bar l_p \gamma_\mu l_r)(\bar d_s \gamma^\mu d_t)$ \\
$Q_{qe}$               & $(\bar q_p \gamma_\mu q_r)(\bar e_s \gamma^\mu e_t)$ \\
$Q_{qu}^{(1)}$         & $(\bar q_p \gamma_\mu q_r)(\bar u_s \gamma^\mu u_t)$ \\ 
$Q_{qu}^{(8)}$         & $(\bar q_p \gamma_\mu T^A q_r)(\bar u_s \gamma^\mu T^A u_t)$ \\ 
$Q_{qd}^{(1)}$ & $(\bar q_p \gamma_\mu q_r)(\bar d_s \gamma^\mu d_t)$ \\
$Q_{qd}^{(8)}$ & $(\bar q_p \gamma_\mu T^A q_r)(\bar d_s \gamma^\mu T^A d_t)$\\
\end{tabular}
\end{minipage}

\vspace{0.25cm}

\begin{minipage}[t]{3.75cm}
\renewcommand{\arraystretch}{1.5}
\begin{tabular}[t]{c|c}
\multicolumn{2}{c}{$8:(\bar LR)(\bar RL)+\hbox{h.c.}$} \\
\hline
$Q_{ledq}$ & $(\bar l_p^j e_r)(\bar d_s q_{tj})$ 
\end{tabular}
\end{minipage}
\begin{minipage}[t]{5.5cm}
\renewcommand{\arraystretch}{1.5}
\begin{tabular}[t]{c|c}
\multicolumn{2}{c}{$8:(\bar LR)(\bar L R)+\hbox{h.c.}$} \\
\hline
$Q_{quqd}^{(1)}$ & $(\bar q_p^j u_r) \epsilon_{jk} (\bar q_s^k d_t)$ \\
$Q_{quqd}^{(8)}$ & $(\bar q_p^j T^A u_r) \epsilon_{jk} (\bar q_s^k T^A d_t)$ \\
$Q_{lequ}^{(1)}$ & $(\bar l_p^j e_r) \epsilon_{jk} (\bar q_s^k u_t)$ \\
$Q_{lequ}^{(3)}$ & $(\bar l_p^j \sigma_{\mu\nu} e_r) \epsilon_{jk} (\bar q_s^k \sigma^{\mu\nu} u_t)$
\end{tabular}
\end{minipage}
\end{center}
\caption{\label{tab:basis}
The independent dimension-six operators built from Standard Model fields which conserve baryon number, as given in 
Ref.~\cite{Grzadkowski:2010es}. The flavour labels $p,r,s,t$ on the $Q$ operators are suppressed on the left hand side of
the tables.}
\end{table}

\end{document}